\documentclass[letter]{aa}

\relax
\usepackage{graphicx}
\usepackage{txfonts}
\usepackage{pdflscape}
\usepackage{color}

\usepackage{amstext}
\usepackage[colorlinks=true,linkcolor=black,citecolor=blue]{hyperref}%

\begin{document} 

\title{Structures Of Dust and gAs (SODA): Constraining the innermost dust properties of II\,Zw96 with JWST observations of H$_2$O and CO}

   \author{I. Garc\'ia-Bernete\inst{1}\fnmsep\thanks{E-mail: igbernete@gmail.com}, M. Pereira-Santaella\inst{2}, E. Gonz\'alez-Alfonso\inst{3}, D. Rigopoulou\inst{1,4}, A. Efstathiou\inst{4}, F.\,R. Donnan\inst{1} and N. Thatte\inst{1}}

  \institute{$^1$Department of Physics, University of Oxford, Keble Road, Oxford OX1 3RH, UK \\
   $^2$ Instituto de F\'isica Fundamental, CSIC, Calle Serrano 123, E-28006, Madrid, Spain\\
   $^3$ Universidad de Alcal\'a, Departamento de F\'isica y Matem\'aticas, Campus Universitario, E-28871, Alcal\'a de Henares, Madrid, Spain\\
   $^4$School of Sciences, European University Cyprus, Diogenes street, Engomi, 1516 Nicosia, Cyprus\\}

\titlerunning{Structures Of Dust and gAs (SODA): constraining the innermost dust properties with H$_2$O and CO}
\authorrunning{Garc\'ia-Bernete et al.}

   \date{Received 27 November 2023 / Accepted 27 December 2023}

  \abstract
   {We analyze \textit{JWST} NIRSpec$+$MIRI/MRS observations of the infrared (IR) gas-phase molecular bands of the most enshrouded source (D1) within the interacting system and luminous IR galaxy II\,Zw\,096. We report the detection of rovibrational lines of H$_2$O $\nu_2$=1-0 ($\sim$5.3-7.2~$\mu$m) and $^{12}$CO $\nu$=1-0 ($\sim$4.45-4.95~$\mu$m) in D1. The CO band shows the R- and P-branches in emission and the spectrum of the H$_2$O band shows the P-branch in emission and the R-branch in absorption. The H$_2$O R-branch in absorption unveils an IR-bright embedded compact source in D1 and the CO broad component features a highly turbulent environment. From both bands, we also identified extended intense star-forming (SF) activity associated with circumnuclear photodissociation regions (PDRs), consistent with the strong emission of the ionised 7.7~$\mu$m polycyclic aromatic hydrocarbon band in this source. By including the 4.5-7.0~$\mu$m continuum information derived from the H$_2$O  and CO analysis, we modelled the IR emission of D1 with a dusty torus and SF component. The torus is very compact (diameter of $\sim$3\,pc at 5~$\mu$m) and characterised by warm dust ($\sim$ 370\,K), giving an IR surface brightness of $\sim$3.6$\times$10$^{8}$\,L$_\sun$/pc$^2$. This result suggests the presence of a dust-obscured active galactic nucleus (AGN) in D1, which has an exceptionally high covering factor that prevents the direct detection of AGN emission. Our results open a new way to investigate the physical conditions of inner dusty tori via modelling the observed IR molecular bands.}

      
   \maketitle


\section{Introduction}
Most galaxies harbour supermassive black holes (SMBHs) in their centres, which indicates they which indicates they may be undergoing an active phase in their evolution (e.g. \citealt{Hickox14}). A significant fraction of cosmic SMBH growth is taking place in a heavily obscured, but intrinsically luminous active galactic nucleus (AGN), which is generally weak or undetected in hard X-rays (e.g. \citealt{Ueda14}). These deeply obscured nuclei are considered to be an important phase of galaxy evolution (e.g. \citealt{Aalto15}), but the dominant power source is still under debate (e.g. \citealt{Veilleux09}). The dusty cores absorb a significant fraction of the intrinsic AGN and starburst (SB) radiation, reprocessing it to emerge at longer wavelengths, peaking in the infrared (IR). In the Local Universe, a significant fraction ($\sim$20-40\%) of luminous IR galaxies (L$_{\rm IR} >$10$^{11}$\,L$_{\odot}$) harbour deeply obscured nuclei (e.g. \citealt{Falstad21,Garcia-Bernete22b,Donnan23a}). Unveiling their inner region is crucial to better understanding growth and evolution processes in luminous IR galaxies.

\begin{figure}
\centering
\par{
\includegraphics[width=7.6cm, clip, trim=0 22 0 0]{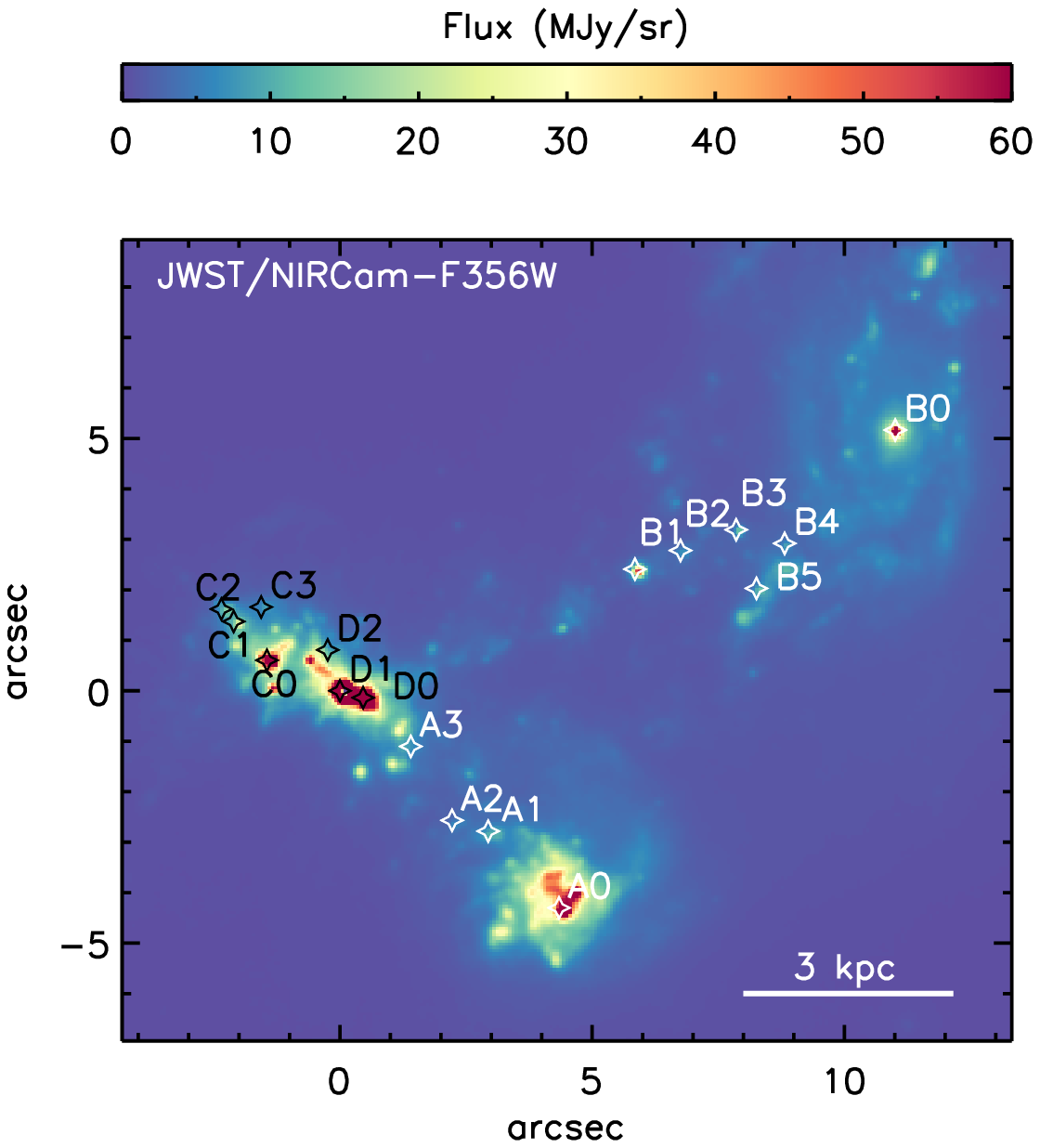}
\par}
\vspace{-5pt}
\caption{IR intensity map of the interacting system II\,Zw\,096. {\textit{JWST}}/NIRCam F356W image (which mainly traces the hot dust and the 3.3\,$\mu$m PAH band). White regions correspond to the A (A0, A1, A2, and A3) and B (B0, B1. B2, B3, B4, and B5) components of the system. Black regions (CO, C1, C2, C3, D0, D1, and D2) are the brightest IR sources in the system. Labeled regions are from \citet{Wu22}. All images are shown in linear colour scale. North is up and east is to the left, offsets are measured relative to D1.}
\label{nircam_f356}
\vspace{-10pt}
\end{figure}

II\,Zw\,096 is a complex interacting system and luminous IR galaxy located at a luminosity distance of 159\,Mpc (z=0.0362). This system consists of four main near-IR sources (A, B, C and D; e.g. \citealt{Goldader97}, see also their Fig. \ref{nircam_f356}). \citet{Goldader97} classified sources A and B as galactic nuclei, whereas they found that the remaining two sources (C and D) are strongly reddened with prominent IR emission. Using {\textit{HST}}/NICMOS 1.6\,$\mu$m imaging observations, D was resolved into two sources: D0 and D1 (\citealt{Inami10}), with the majority of the total IR luminosity coming from D1. \citet{Inami22} demonstrated that D1 is indeed the dominant source of the system in the mid-IR as probed by \textit{James Webb} Space Telescope and Mid-IR Instrument ({\textit{JWST}}/MIRI). These authors also confirmed that the location of D1 coincides with the OH megamasers detected in II\,Zw\,096 (\citealt{Migenes11,Wu22}).  \citet{Migenes11} estimated a lower limit for the enclosed mass of $\sim$10$^9$\,M$_\sun$ in D1, which is consistent with the presence of a SMBH. However, hard X-rays have not been detected from D1 (\citealt{Ricci21}). Given its high hydrogen column density (\citealt{Wu22}), a non-X-ray detection does not rule out a buried AGN.

These previous works have demonstrated the difficulty in providing definitive details regarding the nature of the nuclear-embedded source. II\,Zw\,096 was selected as a {\textit{JWST}} Early Release Science (ERS) target. The unprecedented combination of high angular and spectral resolution (R$\sim$1500-3500) in the 1.0-28.1 $\mu$m range afforded by the {\textit{JWST}} Near-IR Spectrograph (NIRSpec; \citealt{Jakobsen22,Boker22}) and MIRI  ( \citealt{Rieke15, Wells15, Wright15}) are key to investigating deeply embedded sources such as II\,Zw\,096-D1.

{\textit{JWST}} offers a unique advantage to probe warm molecular gas (rovibrational H$_{2}$ lines) and a large number of ionic species covering a wide range of ionisation potentials (IPs), polycyclic aromatic hydrocarbon (PAH) bands, dust features (e.g. silicates), ices (e.g. H$_2$O, CO, CO$_2$), and gas-phase molecular bands (e.g. H$_2$O, CO, CO$_2$, C$_2$H$_2$, and HCN). All of them are excellent tracers to study the AGN-star-formation (SF) connection and the embedded nuclei. In this letter, we report gas-phase rovibrational H$_2$O and CO lines detection in D1, which we have used to characterise its dusty torus properties. To fit the CO and H$_2$O molecular bands, we have taken all the spectroscopic parameters for CO and H$_2$O from the HITRAN2020 database (\citealt{Gordon22}).

\vspace{-10pt}

\section{The IR emission of II\,Zw\,096}
\label{IR emission}

IR imaging and integral field spectroscopy of II\,Zw\,096 was taken using {\textit{JWST}} as part of the Director’s Discretionary Early Release Science (ERS) Program ID:\,1328 (P.I. L. Armus and A. Evans). We refer to  Appendix \ref{reduction} for details on the data reduction. The NIRCam-F356W intensity map of Fig. \ref{nircam_f356} shows a larger level of detail than shown in previously published MIRI maps (\citealt{Inami22}) due to the higher angular resolution at shorter wavelengths. Open stars in Fig. \ref{nircam_f356} correspond to the regions selected by \citet{Wu22} for investigating the CO(3-2) molecular gas of the system. The NIRCam-F356W map mainly traces the hot dust and 3.3\,$\mu$m PAH, with the latter beeing a good star formation (SF) activity indicator in SF galaxies (e.g. \citealt{Rigopoulou99}) and AGN (e.g. \citealt{Castro14}). Thus, the map reveals intense SF activity in those regions located in the C-D zone which also correspond with the brightest IR sources. D1 is the brightest source in the NIRCam-F356W intensity map (Fig. \ref{nircam_f356}).

\vspace{-5pt}

\subsection{Molecular bands of II\,Zw\,096-D1}
\label{IR_emission}
To model the ro-vibrational H$_2$O $\nu_2$=1-0 ($\sim$5.3-7.2~$\mu$m; hereafter H$_2$O band) and $^{12}$CO $\nu$=1-0 ($\sim$4.45-4.95~$\mu$m; hereafter CO band), we removed a baseline continuum in the region of interest (see Appendix \ref{baseline_appendix} for details). Figure \ref{bestmodel} shows a large number of H$_2$O and CO ro-vibrational lines in D1. The H$_2$O band shows the P-branch in emission and the R-branch in absorption as was also found in the galactic Orion BN/KL (\citealt{Alfonso98}). Conversely, both H$_2$O $\nu_2$=1-0 branches are detected in absorption in VV\,114 SW-s2 (\citealt{Alfonso23}). The far-IR and submillimeter (submm) emission of H$_2$O is well documented for luminous IR galaxies (e.g. \citealt{Alfonso10,Alfonso14,Alfonso22,Pereira17}). However, the limited sensitivity provided by previous instruments has prevented the detection of the mid-IR H$_2$O lines in extragalactic sources. The CO band is also detected in D1, but with both the R- and P-branches in emission. This band has been also detected in emission in the outflow region of NGC\,3256 \citep{Pereira23} but appears in absorption toward its AGN nucleus, as well as towards the VV\,114 SW-s2 core (\citealt{Alfonso23}). Earlier studies have investigated the IR CO band using high spectral resolution ground observations in buried galaxy nuclei (e.g. \citealt{Geballe06,Shirahata13,Onishi21,Ohyama23}) and relatively low spectral resolution data using {\textit{Spitzer}} and {\textit{AKARI}} (\citealt{Spoon03,Baba18}). We also detect relatively weak Q-branch C$_2$H$_2$ (13.7$\mu$m) and HCN (14.02$\mu$m) molecular absorption in D1 (Appendix \ref{hcnvib}).

The H$_2$O and CO bands in D1 show several components (see Fig. \ref{bestmodel}). The CO band exhibits a striking broad emission component covering both branches (i.e. the individual broad lines are blended forming a pseudo-continuum emission plateau). This can be explained with the presence of highly turbulent gas (outflow component). Given that the H$_2$O lines are more separated in wavelength than those of the CO, this plateau is not observed in the H$_2$O band. However, H$_2$O broad components are detected in emission in the P-branch and in absorption in the R-branch. The H$_2$O broad P-branch emission might be related to the outflow component, whereas the H$_2$O absorption is produced against a strong IR continuum. The H$_2$O absorption lines trace the dominant luminosity source (dusty torus component). The spectrum of D1 does not show the absorption component of the CO band. This is likely due to the CO broad plateau in emission, which is hiding the CO absorption related to the main luminosity source. 

The CO band clearly also shows narrower emission lines in both branches (SF extended component). While these CO components are present up to the $\sim$P(18) transition, they are not detected for J$>$5 in the R-branch. This indicates a P-R asymmetry in the CO-band (also e.g. \citealt{Alfonso02,Pereira23}). The narrow emission component is detected on top of the highly turbulent H$_2$O component in the P-branch described above. An extra narrow component is needed to explain the observed low-J CO transitions (up to J$\sim$6; cold extended component). The contribution of the cold extended component is not significant for the H$_2$O band. The detected H$_2$O and CO bands demonstrate a great potential for disentangling different components, specifically, four in the case of D1.

\begin{figure*}
\centering
\par{
\includegraphics[width=17.3cm]{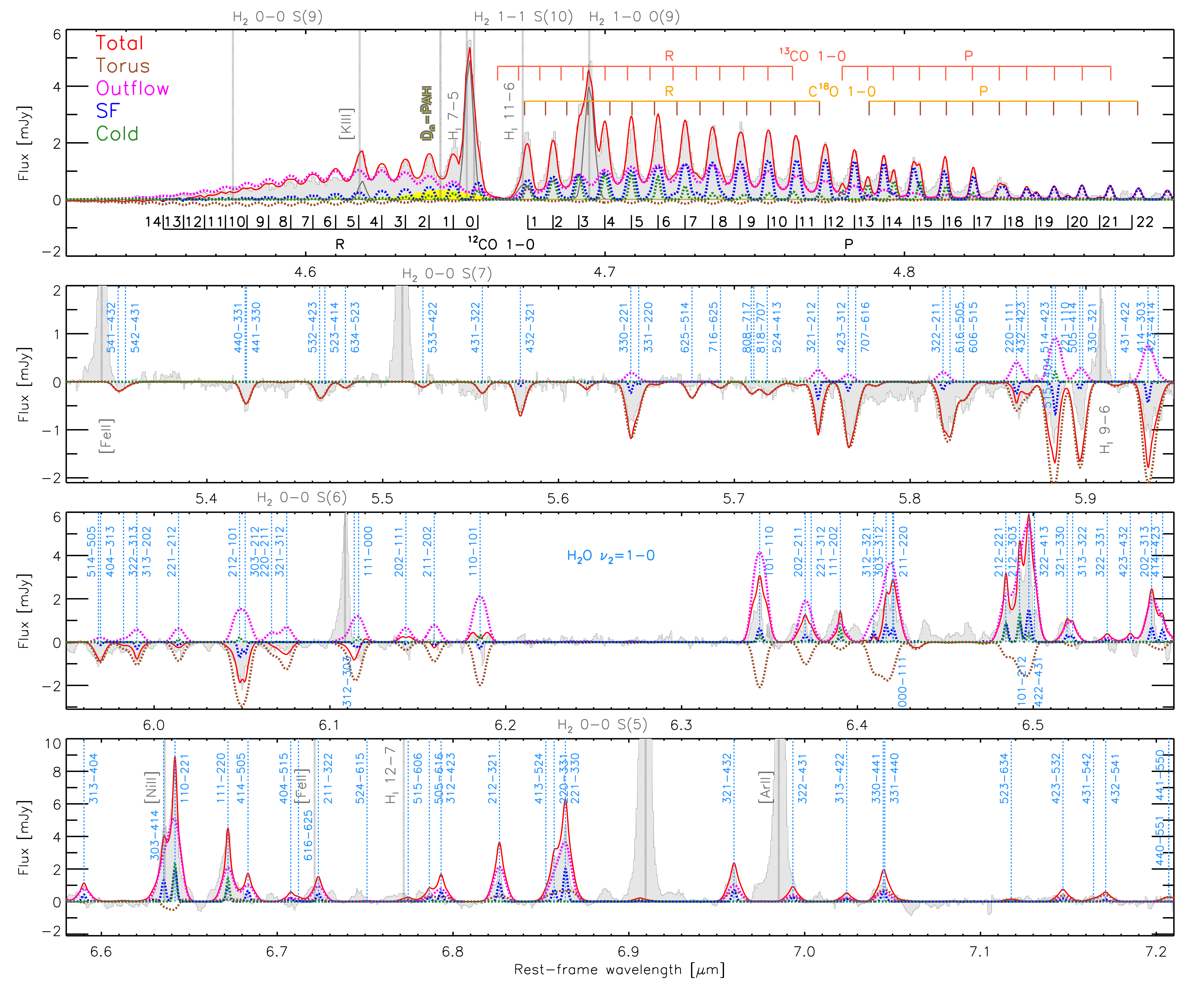}
\par}
\vspace{-5pt}
\caption{Best-fit model for the CO and H$_2$O gas-phase rovibrational bands in II\,Zw\,096-D1. The first panel shows the CO band, while the rest of the panels show the H$_2$O one. The {\textit{JWST}}/NIRSpec and MIRI-MRS rest-frame continuum-subtracted spectra and model fits correspond to the black histograms (filled in gray) and red lines, respectively. We show the model for the torus (brown dashed line), outflow (magenta dashed line), SF-extended (blue dashed line), and cold-extended (green dashed line) components. The vertical lines correspond to the main H recombination, H$_2$, and fine structure lines. Black, red and orange ticks indicate the position of the $^{12}$CO $\nu$=1-0, $^{13}$CO $\nu$=1-0 and C$^{18}$O $\nu$=1-0 rovibrational lines. The blue ticks indicate the H$_2$O $\nu_2$=1-0 rovibational lines that most contribute to the modelled spectrum.}
\vspace{-5pt}
\label{bestmodel}
\end{figure*}

\vspace{-11pt}

\section{H$_2$O and CO modelling}
\label{soda_modelling}
We analysed the rovibrational CO and H$_2$O bands (including the $^{12}$CO $\nu$=1-0, $^{13}$CO $\nu$=1-0, and C$^{18}$O $\nu$=1-0 bands) to characterise the properties of the various components present in the spectrum of II\,Zw\,096-D1. We fit the CO and H$_2$O molecular bands ($\sim$4.5-7.3\,$\mu$m) with a library of tailor-made models for D1. Our analysis also provides valuable predictions of the spectral energy distribution (SED) of the innermost source.

\subsection{Library of tailor-made models and fitting strategy}
\label{thegrid}
Following the methodology presented in \citet{Alfonso14} (also \citealt{Gonzalez-Alfonso21,Alfonso23}), we constructed a set of models for II\,Zw\,096. The model components encompass diverse spherically symmetric gas and dust distributions. Statistical equilibrium computations were done by using collisional rate coefficients for the collisional excitation of CO and H$_2$O with H$_2$ (\citealt{Yang10} and \citealt{Daniel11}). The statistical equilibrium populations of H$_2$O and CO were calculated using non-local and non-LTE radiative transfer comput ationsincluding a careful treatment of blending among lines of the same or different species. In this way, we are able to calculate the fluxes and profiles associated with all relevant spectral lines and the dust continuum. 

For fitting this source, we used four components (as identified in Section \ref{IR_emission}), with different physical properties (T$_{dust}$, T$_{gas}$, column density, gas density, and velocity dispersion). The library of models used for this fit are discussed in Appendix \ref{mygrid}. The components are: 1) the outflow component (turbulent gas) model needed to reproduce the broad plateau observed in both R- and P-branches of the CO band, and in the broad emission lines in the P-branch of H$_2$O band; 2) the dusty torus component, which is mainly responsible for the H$_2$O absorption lines of the R-branch; 3) the SF-extended warm gas component to explain the narrow line emission seen in the CO band, mainly in the P-branch; 4) a colder extended component to reproduce the narrow line excess detected in the low-J transitions of the CO band. We advance that the SF-extended component has an abundance ratio of [CO]/[H$_2$O]$\sim$100 (Section \ref{fitting}), which is compatible with that of the Orion Bar\footnote{In contrast, the submm CO and H$_2$O lines detected with {\textit{Herschel}} have comparable strength in Mrk\,231 (\citealt{Alfonso10}).} (\citealt{Habart10}), the prototypical Galactic photodissociation region (PDR). Therefore, the nature of this component is likely to be related to dense PDRs. Given the rotation pattern observed in the narrow $^{12}$CO $\nu$=1-0 P(13) transition (see Fig. \ref{COP13} in Appendix \ref{resolvedco}), rather than a compact source the SF-extended component is most likely related to an ensemble of PDRs distributed within the beam of {\textit{JWST}} ($\sim$195\,pc). The presence of PDRs is also consistent with the large contribution of the ionised PAH to the mid-IR spectrum of D1. D1 exhibits a prominent 7.7\,$\mu$m PAH band with an elevated 7.7\,$\mu$m/11.3\,$\mu$m PAH ratio ($\sim$50; {\textcolor{blue}{Donnan, in prep}}), compared to values that are usually observed for SF galaxies ($\sim$3.5; \citealt{Bernete22a,Bernete22d}).

To obtain the best-fit model, we minimised the $\chi^2$ in the fit to the entire CO and H$_2$O bands. Given the large number of combinations $\sim$2$\times$10$^9$ and points to be evaluated, the fitting process was performed in three steps. First, we fixed the dusty torus models traced by the H$_2$O absorption lines of the R-branch. Then, we considered all the possible combinations for the remaining components ($\sim$4.7$\times$10$^6$ combinations). We then repeated the fitting process freezing all the components, except the torus component and started the fitting loop again. The fitting and masking of additional emission lines is described in Appendix \ref{mygrid}.

\begin{table}[ht]
\centering
\tiny
\begin{tabular}{lccccccc}
\hline \hline
Comp. &log N$_{\rm H_2}$ & log N$_{\rm CO}$ & log N$_{\rm H_2O}^{\rm orto+para}$& T$_{\rm gas}$ &T$_{\rm dust}$ & V$_{\rm disp}$\\
    & cm$^{-2}$ & cm$^{-2}$ & cm$^{-2}$& K & K & km/s\\
\hline
Torus   &$\cdots$ &$>$18.6&18.6&250&370&140\\
Outflow  & 22.0 &18.5&18.4&250&400&155\\
SF       & 23.1 &19.6&17.7&100&400&35\\
Cold     & 23.1 &19.6&17.7&20&300&20\\
\hline
\end{tabular}                                           
\caption{Main results from the H$_2$O and CO modelling.
Hydrogen column densities are estimated by using the standard value of the CO/H$_2$ abundance of $\sim$3.2$\times$10$^{-4}$ (e.g. \citealt{Bolatto13}). The CO band is in emission in D1, which is not favourable for a definitive estimation of the CO column density of the torus component (in absorption). Thus, we consider the CO column density of the torus a lower limit. We note that the ALMA hydrogen column density of D1 derived in a beam of $\sim$134\,pc$\times$107\,pc is log\,N$_{\rm H}$\,(cm$^{-2})\sim24.1-24.4$ (see Appendix \ref{metallicity}). }
\label{table_results}
\end{table}

\vspace{-23pt}

\subsection{Best-fit model}
\label{fitting}
\begin{figure}
\centering
\par{
\includegraphics[width=6.8cm]{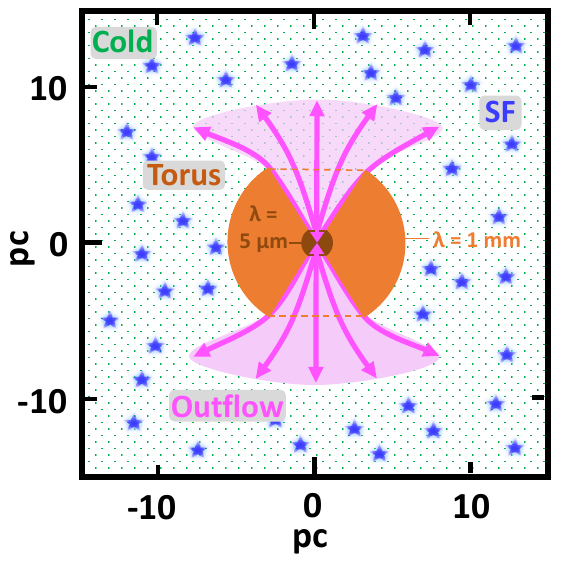}
\par}
\vspace{-10pt}
\caption{The sketch depicts the central region of II\,Zw\,096-D1. The torus, outflow, SF-extended and cold-extended components correspond with the brown, magenta, blue and green colors as in Fig. \ref{bestmodel}.
The orange component corresponds to the torus at 1\,mm derived in Section \ref{modelling_torus}, which is larger than that at $\sim$5\,$\mu$m (brown component). This is expected since submm sizes correspond to the
colder and, thus, more external material within the torus (e.g. \citealt{Burillo21}).}
\vspace{-15pt}
\label{sketch}
\end{figure}

In Fig. \ref{bestmodel}, we compare the observed and best-fit model profiles of the CO and H$_2$O rovibrational lines. The residuals are discussed in Appendix \ref{residuals}. The derived effective size of the nuclear dusty structure is very small ($\sim$1.4\,pc diameter at 5\,$\mu$m) with a warm dust temperature (370$\pm$20K), as the typical torus around AGN (e.g. \citealt{Pier92}). The detection of the C$_2$H$_2$ and HCN absorption bands provide further evidence of the presence of a compact buried nucleus (see Appendix \ref{hcnvib}). The turbulent gas (potential nuclear outflow) has a gas temperature of 250\,K and a relatively high velocity dispersion of $\sim$155\,km/s. The nature of this turbulent region is most likely connected to the inner dusty torus because both exhibit a relatively broad velocity dispersion and similar temperature for the gas. The SF extended component accounts for the observed asymmetry between the CO P- and R- branches. Finally, the cold component, which should be also extended, explains the low energy narrow lines observed in the CO band. This component shows the colder gas temperature ($\sim$20 K). In Table \ref{table_results}, we list the physical parameters of each component. Figure \ref{sketch} shows a sketch summarising all the components.

Our model shows the great potential of the H$_2$O and CO bands to disentangle different components in the inner region of luminous IR galaxies. As mentioned above, we also derive the 4.5-7.0\,$\mu$m dusty continuum associated with the fitted components. Interestingly, the continuum level associated with the torus is about $\sim$40-60\% of the observed flux at 5-7\,$\mu$m. This can provide important insights into the physical conditions of the nuclear warm dust in this source, as shown in continuation.

\vspace{-5pt}

\section{Inner dusty structure modelling: Molecular bands as a tool for constraining the torus properties}
\label{modelling_torus}

Comparing torus models to the observed IR SEDs is an effective method for constraining the properties of the nuclear dusty structure. Torus models with a high covering factor and smooth dust distribution are better suited to reproduce the observed continuum of buried nuclei present in luminous IR galaxies (e.g. \citealt{Levenson07,Efstathiou21,Garcia-Bernete22b}). Here we combine these torus models with the constraints inferred from the models of the H$_2$O and CO bands.
High angular resolution IR data ($\sim$1-20\,$\mu$m) of local AGN (within $\sim$tens of Mpc) isolate most of the nuclear dust emission (inner $\sim$100\,pc; e.g. \citealt{Bernete22c} and references therein). However, the properties of the extreme torus structure in more distant luminous IR systems are poorly understood, even using state-of-the-art IR instrumentation. Their central region consists of a combination of AGN and an intense SF activity component (e.g. \citealt{Paredes15,Herrero-Illana17,Efstathiou21}). Thus, best-fit solutions might be subject to multi-component fitting degeneracies.

\begin{figure*}
\centering
\par{
\includegraphics[width=9.0cm]{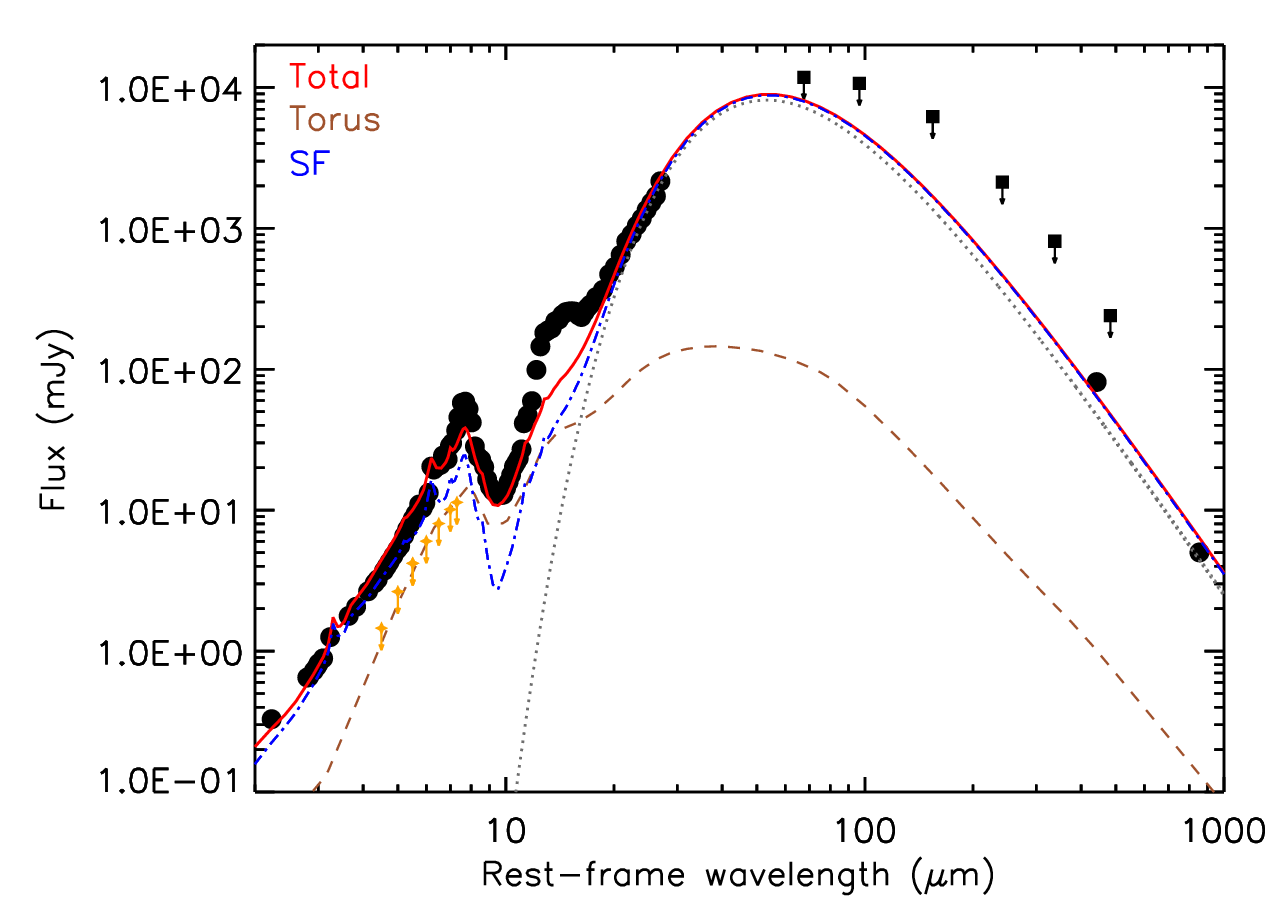}
\includegraphics[width=8.0cm, clip, trim=0 4 0 0]{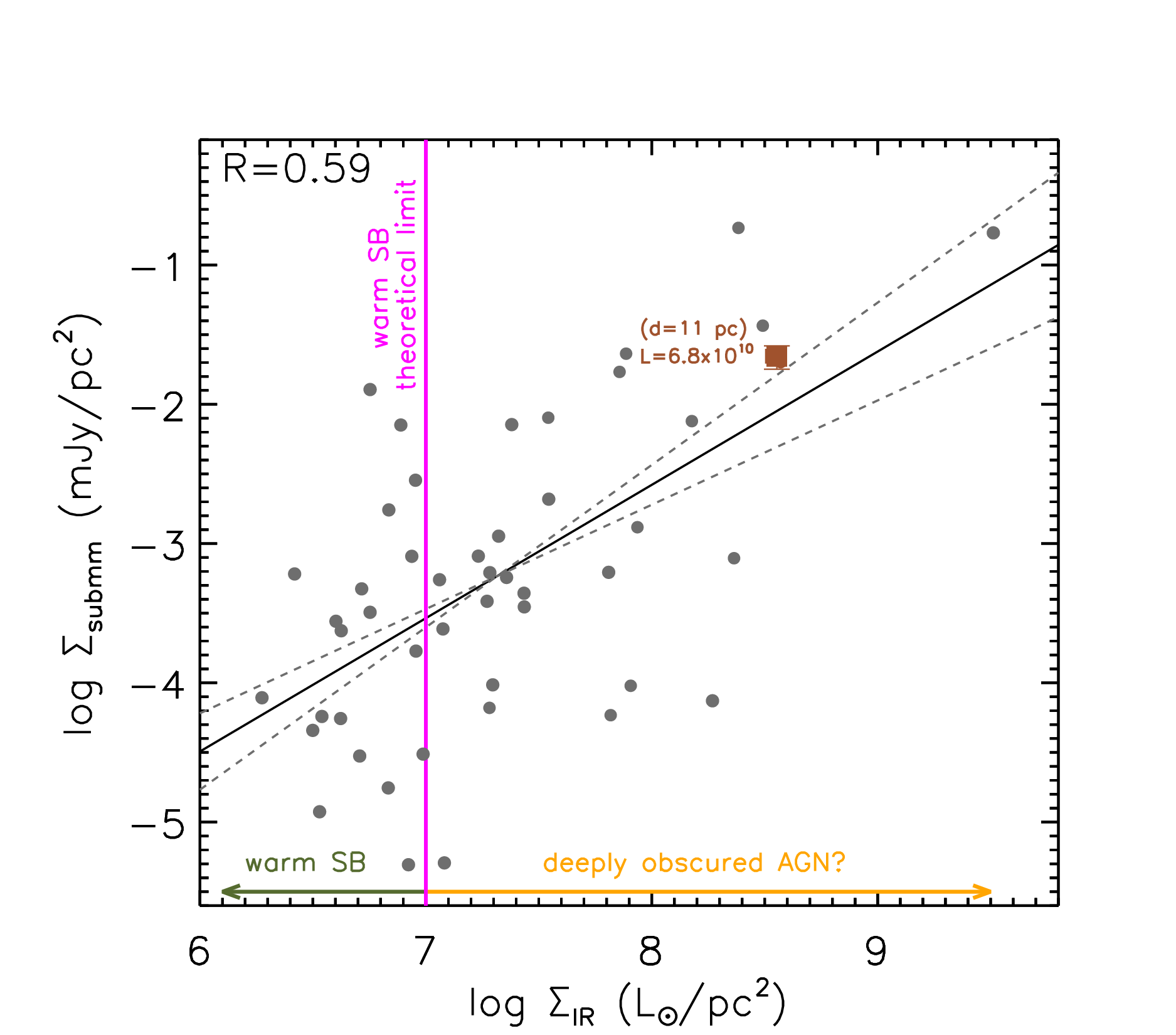}
\par}
\caption{Left panel: IR-to-submm SED of II\,Zw\,096-D1 (black circles: {\textit{JWST}} spectrum and {\textit{ALMA}} continuum photometric data; black squares: {\textit{Herschel}} fluxes from \citealt{chu17}). Orange stars correspond to the derived IR continuum from the observed H$_2$O and CO bands. Dashed brown and dot-dashed blue lines correspond to the torus and SF-extended components. The dotted grey curve represents a grey body of 57$\pm$4\,K. The solid red line corresponds to the total model. Right panel: Submm vs. IR surface brightness. The magenta vertical line represents the theoretical limit for a warm SB (\citealt{Thompson05}). Brown square and grey circles correspond to the surface brightness of D1 found in this work and those of IR galaxies from \citet{Falstad21}. Grey lines correspond to the best linear fit (R=0.59; log ($\Sigma_{\rm submm}$)=log ($\Sigma_{\rm IR}$)$\times$ (0.96$\pm$0.21) - 10.24$\pm$1.53).}
\label{torus}
\end{figure*}

Here, we use the radiative transfer CYprus models for Galaxies and their NUclear Spectra (CYGNUS) project (\citealt{Efstathiou21}). We refer to this work and the references therein for further details on how the models are constructed. To fit the SED of D1, we used a Markov chain Monte Carlo (MCMC) fitting code (SED analysis through Markov chains; \citealt{Johnson13}), which includes Bayesian statistics. 

To build the 1-850\,$\mu$m SED of D1, we used the entire NIRSpec and MRS spectra (resampled to $\sim$100 points), high angular resolution {\textit{ALMA}} archival data (Appendix \ref{reduction}), and lower angular resolution far-IR fluxes from {\textit{Herschel}}. We measured the central fluxes in the {\textit{ALMA}} data using a point-source extraction, and compiled the PACS and SPIRE point-source extracted fluxes from \citet{chu17}. However, given the large physical area traced by the {\textit{Herschel}} data (ranging from $\sim$5.3 to 34.7\arcsec; $\sim$3.8 to 25\,kpc), we considered to set them as upper limits in the fit. 

As expected in luminous IR galaxies hosting AGN, including dusty torus and SF components is required to reproduce the observed SED of D1. We include an additional grey body of 57$\pm$4\,K and $\beta$=1.8 (following the same method as in \citealt{Kovacs10}). This component corresponds to the relatively cold dust associated with the SF component. Thus, we plot both components together in the left panel of Fig. \ref{torus} (blue dashed line). 

The best-fit torus model (without using any prior information from the bands) overestimated the intrinsic continuum in the range 4.9-7.0\,$\mu$m relative to the continuum derived from the H$_2$O and CO bands. The resulting fit has a high fractional contribution of the torus to the SED at 7~$\mu$m (F$_{7~\mu m}^{torus}\sim$70\%). We can discard this scenario given that this strong central component would produce very deep absorption bands that are not observed in D1. The torus continuum level is indeed constrained by the observed H$_2$O absorption band. This provides an observational limit to the contribution of the torus. 

We repeated the fitting process but including the 4.9-7.0\,$\mu$m SED derived from the H$_2$O and CO analysis (Fig. \ref{torus}, left). The fractional contribution of the torus to the SED is lower (F$_{7~\mu m}^{torus}\sim$40\%), and the derived torus bolometric luminosity is about nine times smaller. Interestingly, the 4.9-7.0\,$\mu$m continuum of the SF component is also consistent with that reported in Section \ref{fitting}. The derived torus diameter is $\sim$3\,pc and $\sim$11\,pc at 5\,$\mu$m and 1\,mm, respectively (see Appendix \ref{torusimages}). The size of the dusty structure at 5\,$\mu$m derived from the torus model is roughly twice as large as that determined by the molecular bands fit. This difference arises from the torus models accounting for temperature gradients across the dust, whereas the model of molecular bands includes a single temperature blackbody. Hereafter, we use the size derived from torus models at 5\,$\mu$m and 1\,mm.

The resulting fit in Fig. \ref{torus} (left) underestimates the continuum emission at $\sim$12-15$\mu$m. We further investigate the library of torus models, finding that models that reproduce 9.7\,$\mu$m silicate absorption band as well as the emission at $\sim$12-15$\mu$m overestimate the continuum emission at $\sim$7$\mu$m. One possible explanation for the challenges in reproducing the silicate bands of this source could be related to its sub-solar metallicity\footnote{The extreme PAH flux ratios found in D1 might be related to the low metallicy of this source (see e.g. \citealt{Li20} for a review).} (Z$\sim$0.6-0.8\,Z$\sun$) and its relatively high gas-to-dust ratio (70-135; Appendix \ref{metallicity}). This suggests that a hard radiation field is present in D1, which might be affecting its dust properties (\citealt{remy14}). Torus models generally assume the standard Galactic mix of 53\% silicates and 47\% graphite, which might be different in lower metallicity systems. A different composition of the dust grains can significantly affect the silicate absorption band (e.g. \citealt{Tsuchikawa21,Tsuchikawa22}). However, investigating the mineralogical properties of dust grains is beyond the scope of this work.

\vspace{-5pt}

\section{Alternative ways of looking for AGN in D1}
\label{myagn}
The superb sensitivity of the {\textit{JWST}} allows us to search for faint high IP lines (see Appendix \ref{lines_appendix}). [Ca\,IV]$\lambda3.21\mu$m and [Mg\,IV]$\lambda4.49\mu$m transitions are detected with a signal-to-noise ratio (S/N) of $>$3 (IP: 50.9 and 80.1\,eV, respectively). The IP of [Ca\,IV] is not enough to firmly establish nuclear activity. Similarly, [Mg\,V] can be detected in non-AGN dominated regions of galaxies with publicly available {\textit{JWST}} data, with unclear origin (\textcolor{blue}{Pereira-Santaella, in prep}). An emission line at 1.43\,$\mu$m coincides with the expected position of [Si\,X] (IP: 351.1\,eV), but [Si\,VI] is not detected, which is typically brighter than [Si\,X] in AGN (1.8 times on average; \citealt{rodriguez-ardila11}). This indicates that the emission line observed at 1.43$\mu$m is not likely to be related to [Si\,X].

Although AGN emission is not clearly detected, there is evidence for the presence of an accreting SMBH in D1. The diameter of the torus is very compact ($\sim$3 and $\sim$11\,pc at 5\,$\mu$m and 1\,mm, respectively) with a very warm dust temperature ($\sim$370 K). Following the same argument as that in \citet{Downes07}, we estimate that nearly $\sim$7$\times$10$^{5}$ O stars need to be present in only $\sim$11\,pc (a similar luminosity of M\,82 but from a few thousand times smaller volume). The density of the required O stars would be consistent with the value derived from the buried AGN present in Arp\,220 (west nucleus; \citealt{Downes07}). The dusty structure bolometric luminosity is $\sim$6.8$\times$10$^{10}$\,L$_\sun$ and its IR surface brightness\footnote{$\Sigma_{\rm IR}$=(L$_{\rm IR}$/2$\pi$r$^2$) and $\Sigma_{\rm submm}$=(F$_{\rm submm}$/2$\pi$r$^2$) are calculated using anisotropic IR luminosity and 1\,mm flux from the best-fit torus model and a diameter of 11\,pc.} ($\Sigma_{\rm IR}$) is $\sim$3.6$\times$10$^{8}$\,L$_\sun$/pc$^2$. We plot in Fig. \ref{torus} (right) the relationship between the submm surface brightness and that of the IR for luminous IR galaxies (\citealt{Falstad21}),  finding that both surface brightness broadly correlate (R=0.59) and the values derived in this work for D1 follow this relation. The elevated surface brightness found for this source is consistent with a buried AGN.

Reconciling the IR surface brightness of D1 with a compact SB is challenging. For instance, \citet{Rico-Villas22} found a proto-super star cluster in NGC\,253, which is extremely buried (N$_{\rm H}\sim$10$^{25}$\,cm$^{-2}$), with a IR surface brightness of  $\sim$1.3$\times$10$^{7}$\,L$_\sun$/pc$^2$, which is $\sim$30 times smaller than that of D1. In addition, from the theoretical point of view, there is a limit of $\sim$10$^{7}$\,L$_\sun$/pc$^2$ for a maximal SB (e.g. \citealt{Thompson05,Pereira-Santaella21}). Although we cannot rule out the possibility of an extreme top-heavy initial mass function (IMF) SB, this is unlikely since these systems are not stable on short timescales of few Myr (see e.g. \citealt{Pereira-Santaella21}). 

\vspace{-10pt}

\section{Summary and conclusions}
\label{conclusions}
We present a {\textit{JWST}\,NIRSpec$+$MIRI/MRS study of the IR gas-phase H$_2$O ($\sim$5.3-7.2~$\mu$m) and CO ($\sim$4.45-4.95~$\mu$m) rovibrational bands from the most enshrouded source (D1) in the interacting system, luminous IR galaxy II\,Zw\,096. Our main results are as follows.

\begin{enumerate}
\item  We identified several components in the rovibrational H$_2$O and CO bands in D1. By comparing the observed molecular rovibrational bands with radiative transfer models, we find that D1 is a complex structure featuring an IR-bright embedded dusty core (i.e. torus), which is most likely connected to a highly turbulent environment. We also detected widespread PDR emission, which is consistent with the strong emission of the ionised 7.7$\mu$m PAH band in this source. \\

\item  We modelled the IR emission of D1 using dusty torus and SF components, including the constraints at 4.5-7.0\,$\mu$m SED inferred from the H$_2$O and CO analysis. The size of the dusty structure is very compact ($\sim$3 and $\sim$11\,pc at 5\,$\mu$m and 1\,mm, respectively) with a warm dust temperature ($\sim$370 K). The dusty structure bolometric luminosity is $\sim$6.8$\times$10$^{10}$\,L$_\sun$ and its IR surface brightness is $\sim$3.6$\times$10$^{8}$\,L$_\sun$/pc$^2$, suggesting the presence of a compact and dust-obscured AGN in D1. The exceptionally high covering factor of the dusty structure inhibits the direct detection of AGN emission.

\end{enumerate}

In summary, this pilot study employing {\textit{JWST}}/NIRSpec$+$MRS spectroscopy demonstrates the potential of the H$_2$O and CO rovibrational bands for gaining insights into the inner regions of luminous IR galaxies, especially when spatially unresolved. 
New observations of luminous IR galaxies with {\textit{JWST}} will allow for subsequent detailed studies,  which will be crucial for enhancing the statistical significance of the findings presented here.

\begin{acknowledgements}
The authors acknowledge the DD-ERS teams for developing their observing program with a zero--exclusive--access period. IGB and DR acknowledge support from STFC through grants ST/S000488/1 and ST/W000903/1. EG-A acknowledges grants PID2019-105552RB-C4 and PID2022-137779OB-C41 funded by the Spanish MCIN/AEI/10.13039/501100011033. MPS acknowledges support from grant RYC2021-033094-I funded by MCIN/AEI/10.13039/501100011033 and the EU NextGenerationEU/PRTR. The authors are extremely grateful to Susana Guadix-Montero for helping with the sketch, and to the {\textit{JWST}} helpdesk for their support. Finally, we thank the anonymous referee for their useful comments.

\vspace{-15pt}

\end{acknowledgements}

\begin{appendix}

\section{Data reduction}
\label{reduction}

\subsection{{\textit{JWST}}}
We retrieved near-IR to mid-IR (2.87-5.27~$\mu$m) data observed using  integral-field spectrographs MIRI MRS with a spectral resolution of R$\sim$3700--1300 (\citealt{Labiano21}) and NIRSpec with the grating-filter pairs G140H (0.97–1.89~$\mu$m), G235H (1.66–3.17~$\mu$m) and G395H (2.87–5.27~$\mu$m) with R$\sim$2700 (\citealt{Jakobsen22,Boker22}). We primarily followed the standard MRS pipeline procedure (e.g. \citealt{Labiano16} and references therein) and the same configuration of the pipeline stages described in \citet{Bernete22d} and \citet{Pereira22} to reduce the data. Some hot and cold pixels are not identified by the current pipeline, so we added some extra steps as described in \citet{Pereira23} and \citet{Bernete23} for NIRSpec and MRS, respectively.

We extracted D1 spectrum by simultaneously applying 2D Gaussian models to both D1 and D0 to correctly deblend them at longer wavelengths. To do so, we employed observations of calibration point sources (MRS HD-163466 and IRAS,05248$-$7007, Programme IDs 1050 and 1049) to measure the width and position angle of a 2D Gaussian for each spectral channel. To obtain the point source flux we used the models of the calibration PSF stars from \citet{Bohlin20}, which is equivalent to applying aperture correction factors. We refer the reader to \citet{Bernete23} for further details. We also note that D1, the primary focus of this study, is several times brighter than D0 (3-5 times in the mid-IR; \citealt{Inami22}).

We also downloaded the fully reduced and calibrated NIRCam F356W ($\lambda_c=$ 3.563~$\mu$m) observations to produce a high angular resolution IR intensity map for the system (see Fig. \ref{nircam_f356}).

\subsection{{\textit{ALMA}}}
II\,Zw\,096 was observed in Bands 7 and 9 using the 12m {\textit{ALMA}} array. HCN-vibrational (4--3) 356.256\,GHz transition and 350.9\,GHz continuum were observed in Band 7 and the 701.5\,GHz continuum in Band 9. These data were obtained as part of programmes 2011.0.00612.S, 2012.1.01022.S, and 2017.1.01235.S (PI. S. Stierwalt, and L. Barcos-Mu\~{n}oz). For the Band 7 data, we combined two datasets using
extended and compact array configurations to increase the uv-plane coverage. We used the standard {\textit{ALMA}} pipeline (CASA v6.2.1; \citealt{McMullin07}) to calibrate and clean the data. The synthesised beam full-width half-maximum (FWHM) is 0\farcs19$\times$0\farcs16 and the maximum recoverable scale is 5\farcs7. We also used the standard {\textit{ALMA}} scripts to calibrate the Band 9 data. Because of the higher atmospheric variability at Band 9, the default phase calibration was not enough to correct the data. Thus, we self-calibrated the phase of the visibilities using the peak of the CO(6--5) emission. The synthesised beam FWHM is 0\farcs28$\times$0\farcs24 and the maximum recoverable scale is 2\farcs5. 

\section{Baseline}
\label{baseline_appendix}
We subtract a baseline the continuum emission and the broad PAH emission contribution to the II\,Zw\,096-D1 spectrum. To do so, we have consistently used the same method for fitting the baseline of the CO and H$_2$O bands. First, we fit the PAH profile using a modified version of PAHFIT to work with the higher spectral resolution NIRSpec and MRS data (\citealt{Donnan23b,Donnan23a}). The mid-IR modelling of this source will be presented in \textcolor{blue}{Donnan (in prep)}. Using the PAH-subtracted spectrum, we then fit a cubic polynomial function masking all the remaining features present in the spectrum (see Fig. \ref{baseline}).

\begin{figure}
\centering
\par{
\includegraphics[width=9.3cm]{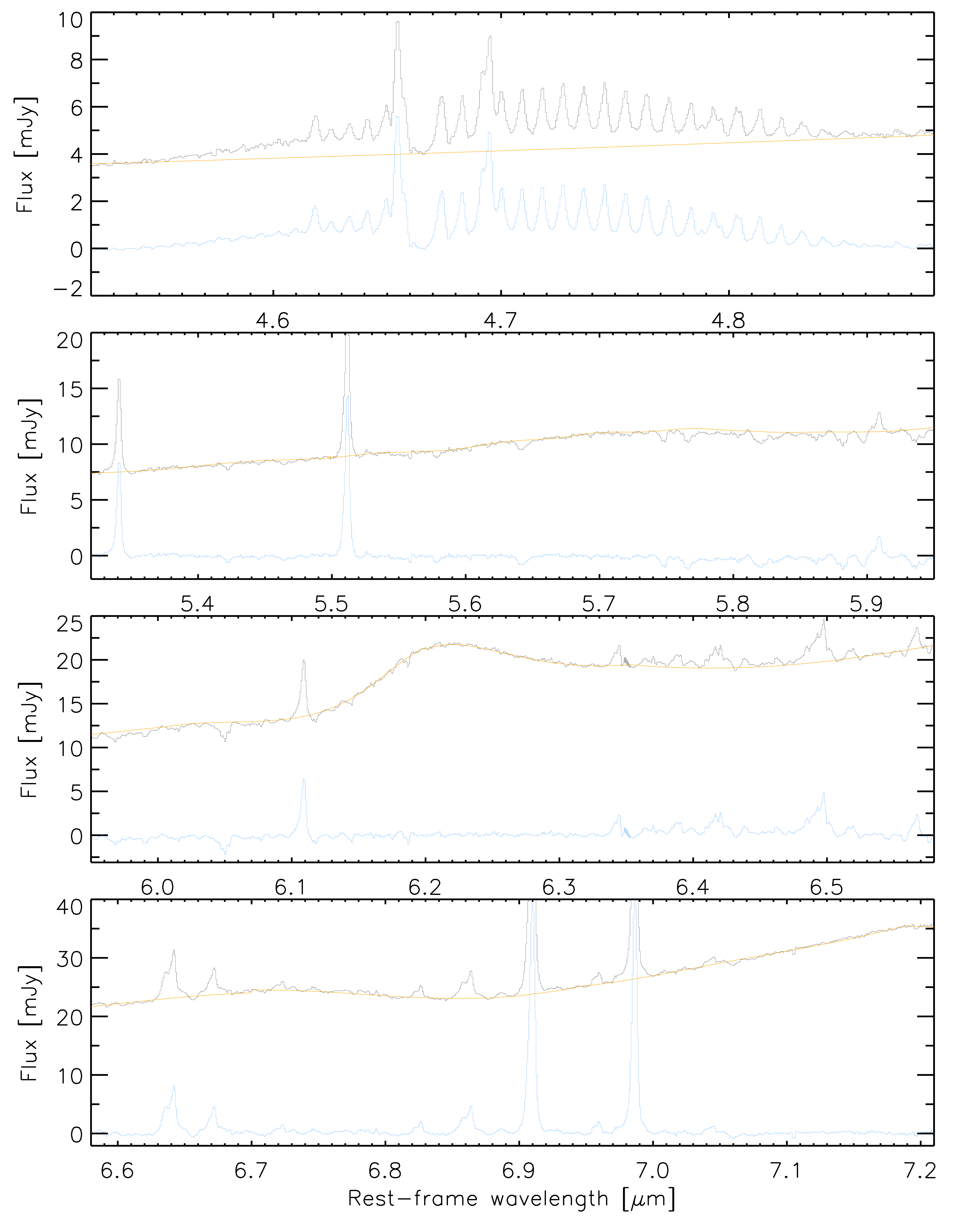}
\par}
\caption{Fitted baseline of II\,Zw\,096-D1. The {\textit{JWST}}/NIRSpec and MRS rest-frame spectra and model fit (i.e. baseline$+$PAH features) correspond to the solid and orange lines. The baseline subtracted spectra is represented by a solid blue line.}
\label{baseline}
\end{figure}

\section{C$_2$H$_2$, HCN and submm HCN-vib (4-3) detection}
\label{hcnvib}
\begin{figure}[ht]
\centering
\par{
\includegraphics[width=8.3cm]{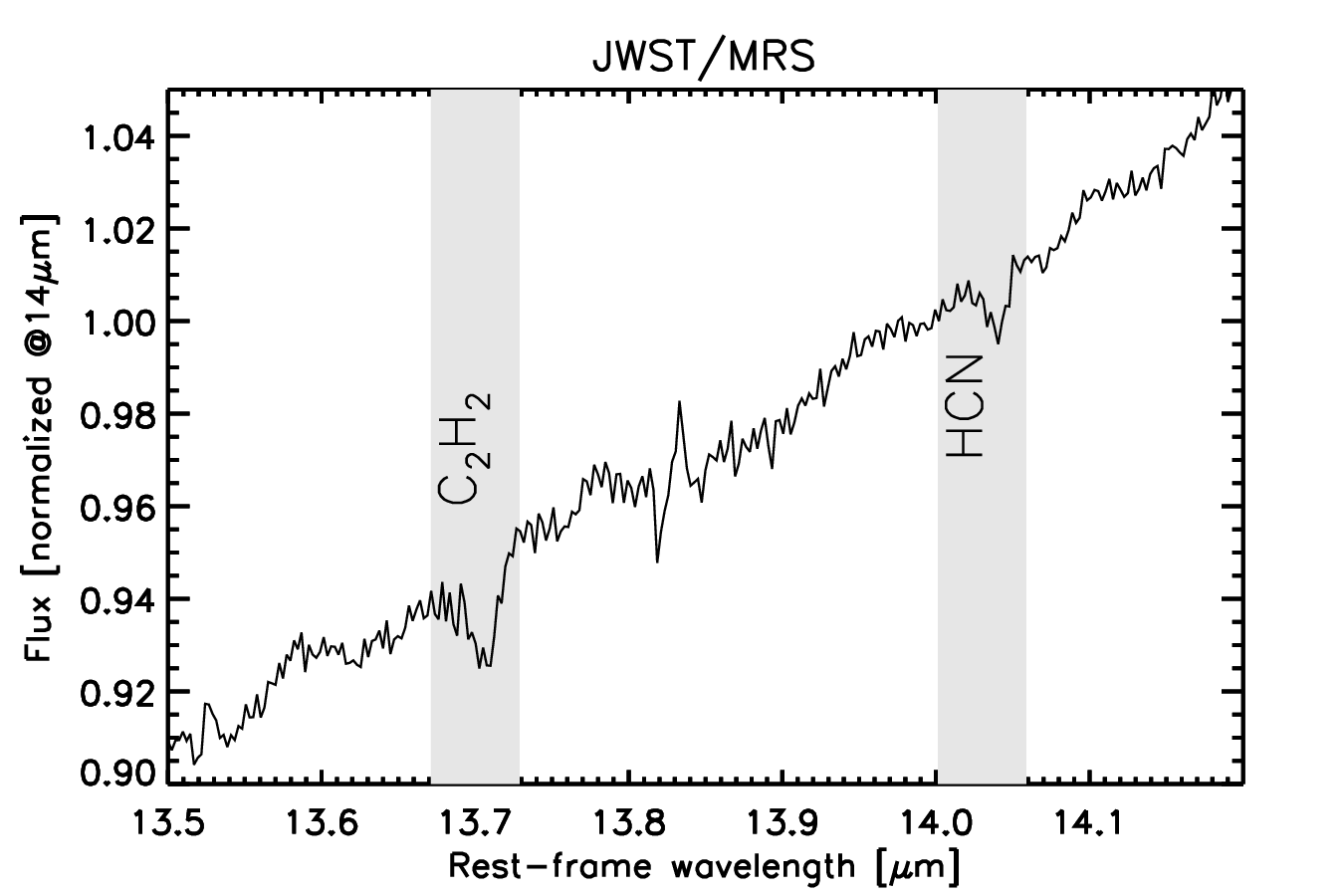}
\includegraphics[width=8.3cm]{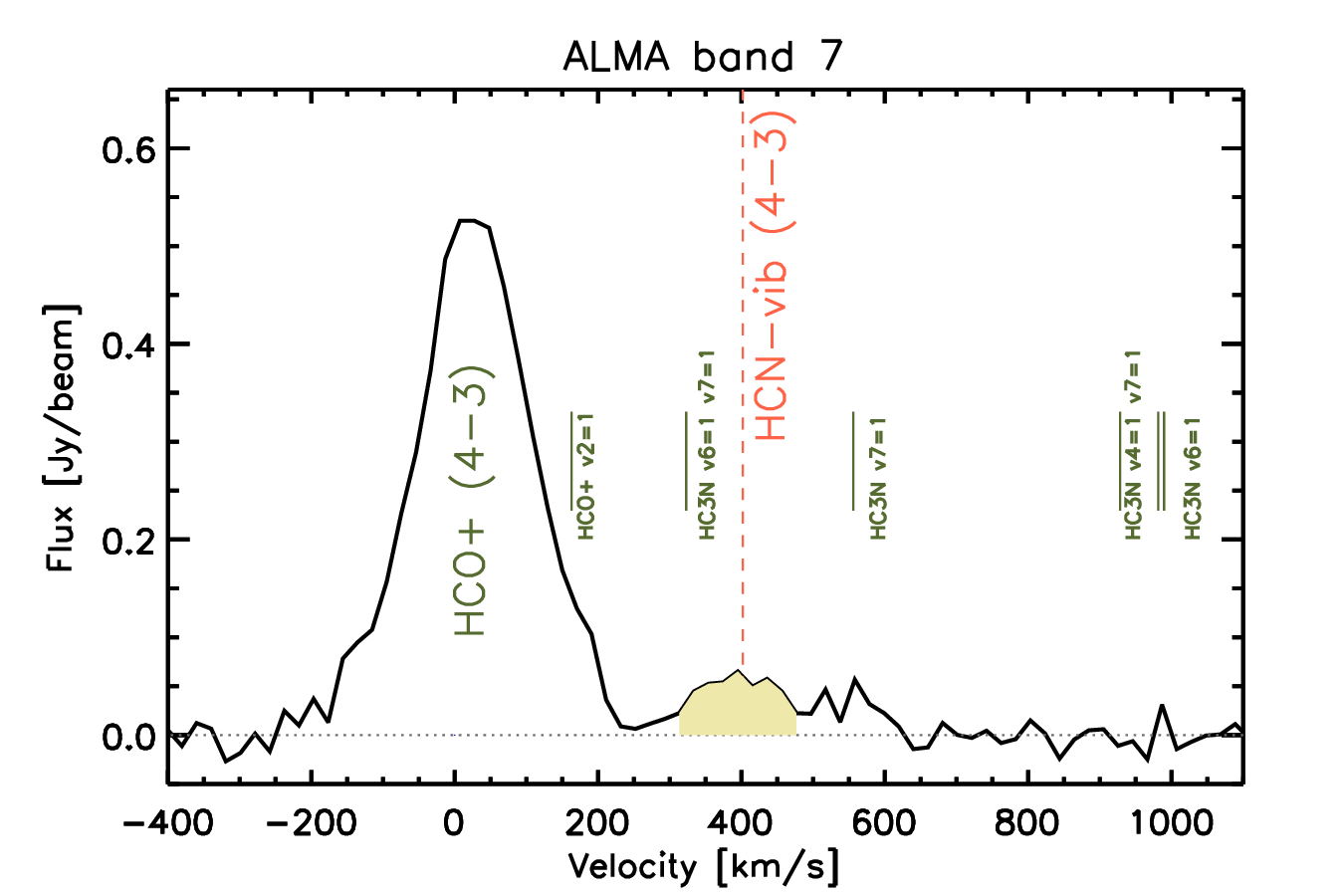}
\par}
\caption{HCN emission of D1. Top: {\textit{JWST}}/MRS spectrum showing the absorption bands of C$_2$H$_2$ and HCN. The spectrum is normalised at 14\,$\mu$m. Bottom: {\textit{ALMA}} submm spectrum showing the HCO$+$ and HCN-vib emission.}
\label{hcn}
\end{figure}
We report the detection of HCN and C$_2$H$_2$ IR bands in II\,Zw\,096-D1, which were non-detected in the previous {\textit{Spitzer}}/IRS observations. The top panel of Fig. \ref{hcn} shows the detected C$_2$H$_2$ and HCN absorption bands in the {\textit{JWST}} spectrum. These are considered excellent tracers of buried nuclei (e.g. \citealt{Lahuis07}). The HCN 14\,$\mu$m absorption band is relatively weak in II\,Zw\,096-D1. The presence of the HCN in absorption suggest that the vibrationally pumped submm HCN (HCN-vibrational) emission in this nucleus might be present, given that this feature is responsible for populating the levels originating the HCN-vibrational emission (e.g. \citealt{Alfonso19}). Indeed, the bottom panel of Fig. \ref{hcn} shows that the HCN-vibrational (4--3) 356.256\,GHz is clearly detected in D1 (see Appendix \ref{reduction} for details on the ALMA data reduction). This provides further evidence of the compact and deeply embedded nature of II\,Zw\,096-D1 (e.g. \citealt{Sakamoto10,Aalto15}). 

\section{Molecular grid of models and fitting of emission lines}
\label{mygrid}
The model grids are classified into four groups: a) torus, b) outflow, c) SF extended component and d) cold extended component. The torus models (396) have N$_{CO}$=1-4$\times$10$^{18}$ cm$^{-2}$, T$_{dust}$=300-400\,K and $^{12}$CO/$^{13}$CO=50-70. The outflow models (243) have CO density of 2-8$\times$10$^6$ cm$^{-3}$, T$_{dust}$=300-400\,K, $^{12}$CO/$^{13}$CO=50-70 and velocity dispersion ranging from 200 to 240\,km/s. The SF extended component models (243) have N$_{CO}$=1-4$\times$10$^{19}$ cm$^{-2}$, T$_{dust}$=300-400\,K, $^{12}$CO/$^{13}$CO=50-70 and a range of covering factors to produce the P-R asymmetry. The cold extended component models (81) is represented by a cold component with T$_{gas}$=20-40\,K, illuminated by an IR source with T$_{dust}$=300-400\,K and $^{12}$CO/$^{13}$CO=50-70. Each of these 4 groups covers a regular grid in the free parameters. We sample the torus model temperature with $\vartriangle$T$_{dust}$=10\,K to estimate the uncertainty associated to this parameter that is crucial for constraining the torus model.

To model the ro-vibrational H$_2$O and CO bands, we also fit and masked additional emission lines within the spectral range of the molecular bands. While we masked additional emission lines within the H$_2$O band ([Fe\,II]\,$\lambda5.34\mu m$, H$_2$\,0-0\,S(7)\,$\lambda5.51\mu m$, HI 9-6\,$\lambda5.91\mu m$, H$_2$\,0-0\,S(6)\,$\lambda6.11\mu m$, [Ni\,II]\,$\lambda6.64\mu m$, [Fe\,II]\,$\lambda6.72\mu m$, HI 12-7\,$\lambda6.77\mu m$, H$_2$\,0-0\,S(5)\,$\lambda6.91\mu m$, and [Ar\,II]\,$\lambda6.99\mu m$), this is not doable for the CO band for which the individual lines are not as separated in wavelength as in the case of H$_2$O. Thus, we also included emission from H$_2$\,0-0\,S(9)\,$\lambda4.58\mu m$, HI 7-5\,$\lambda4.65\mu m$, HI 11-6\,$\lambda4.67\mu m$, [K\,III]\,$\lambda4.62\mu m$ and D$_n$-PAH (4.65\,$\mu$m). We fixed the FWHM of the PAH feature and the flux ratio of the HI recombination (7-5 and 11-6) and [K\,III] lines to the values derived in \citet{Pereira23}.

\section{Spatially resolved mid-IR CO}
\label{resolvedco}
Figure \ref{COP13} shows the spatially resolved mid-IR $^{12}$CO $\nu$=1-0 P(13) emission in II\,Zw\,096-D1. The narrow component of the CO P(13) emission line is not contaminated by the $^{13}$CO and C$^{18}$O (Fig. \ref{bestmodel}). Although the intensity map is centrally concentrated (top panel of Fig. \ref{COP13}), the size fitted with a Gaussian profile is larger than that measured in a calibration point source (NIRSpec TYC\,4433-1800-1, Program ID 1128, PI: N. Luetzgendorf). Furthermore, the velocity field of the $^{12}$CO 1-0 P(13) emission feature indicates rotation of the SF extended component (bottom panel of Fig. \ref{COP13}). We note that the comparison between the PSF star and D1 has been done in the cubes oriented in the instrument integral field unit plane (IFUALIGN) to mitigate PSF orientation effects. This is consistent with the extended PDR component found in Section \ref{soda_modelling}.

\begin{figure}
\centering
\par{
\includegraphics[width=6.0cm]{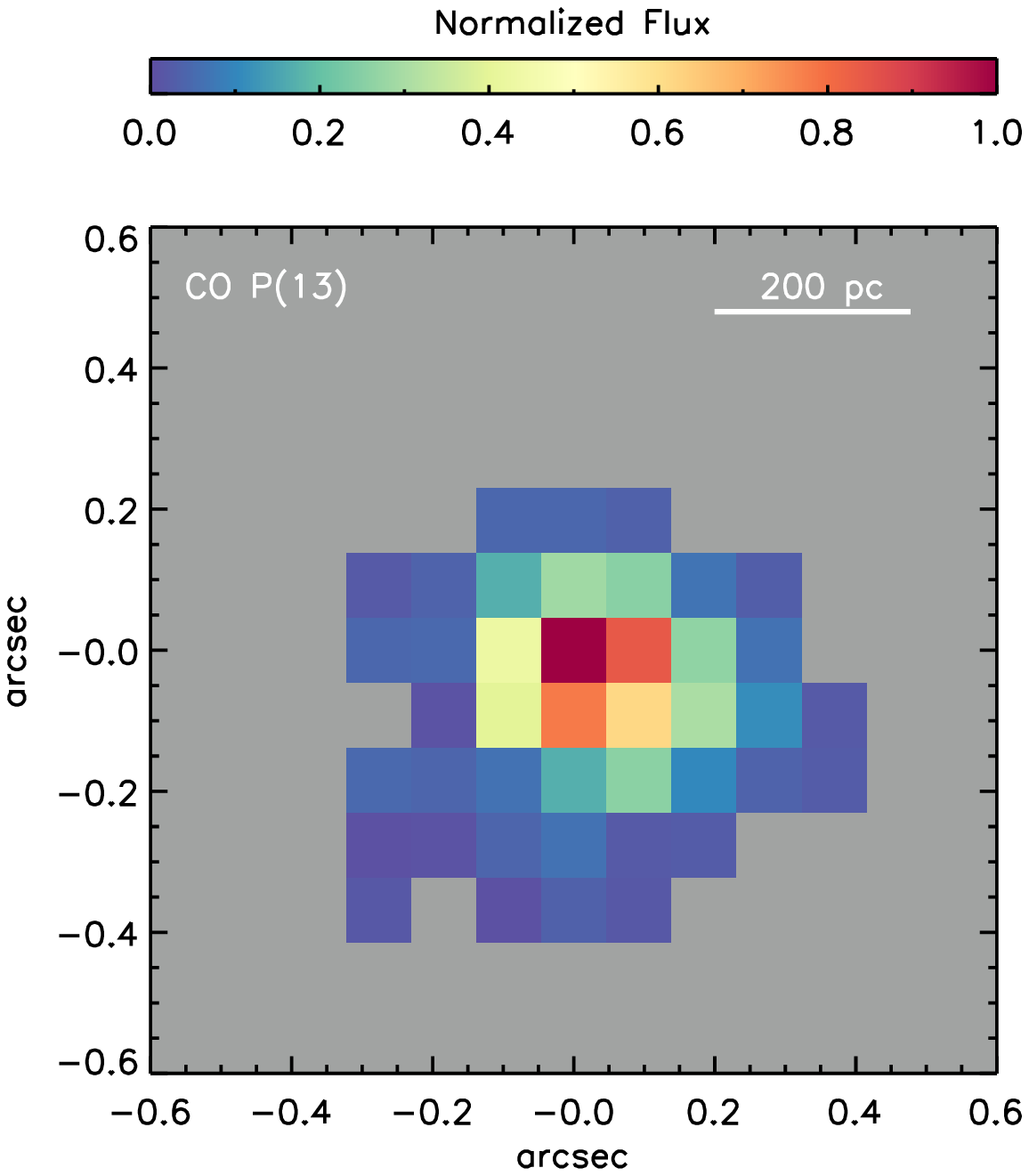}
\includegraphics[width=6.0cm]{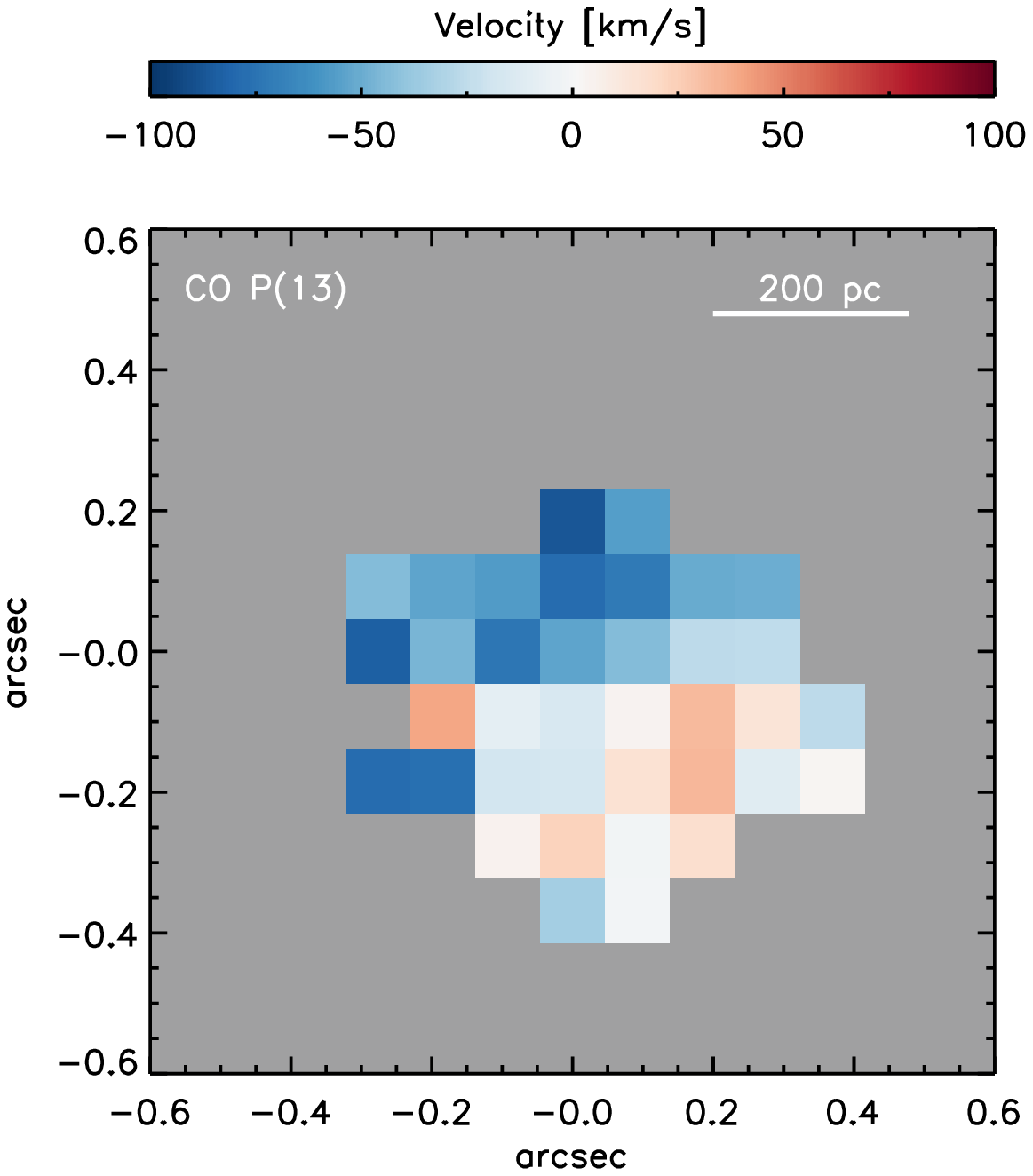}
\par}
\caption{{\textit{JWST}}/NIRSpec $^{12}$CO 1-0 P(13) emission map derived using a local continua (see text). Top panel: Intensity map. Bottom panel: Velocity map.}
\label{COP13}
\end{figure}

\section{Residuals of the best-fit model of the molecular bands}
\label{residuals}

\begin{figure*}
\centering
\par{
\includegraphics[width=16.3cm]{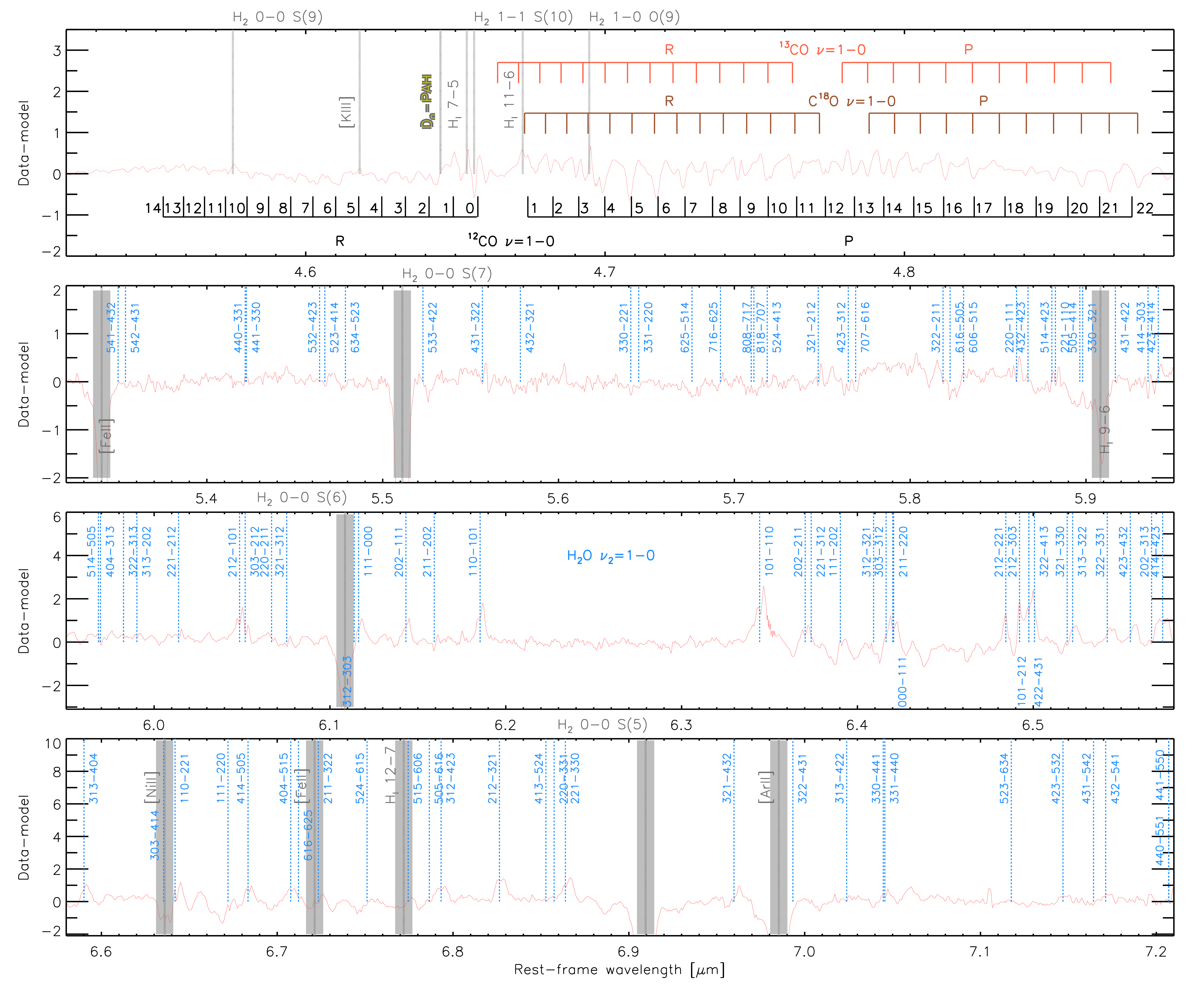}
\par}
\caption{Residuals of the best-fit model for the CO and H$_2$O gas-phase rovibrational bands in II\,Zw\,096-D1.}
\label{bestmodel_residuals}
\end{figure*}
Figure \ref{bestmodel_residuals} shows the residuals of the best-fit model. The residuals at low-J lines (J$<$3) of the $^{12}$CO $\nu$=1-0 band might be related with galactic diffuse or translucent clouds with low gas density and excitation temperature. It is probable that these clouds are also contributing to the H$_2$O $\nu_2$=1-0 band, which clearly shows residuals in the ro-vibrational lines connected to the $\nu_2$=0 low-excitation rotational levels (e.g. 1$_{10}$-1$_{01}$). The enhancement of gas-phase H$_2$O ground vibrational states could result from far-ultraviolet (FUV) photodesorption originating from grains coated with ice in translucent clouds, or in the envelopes of molecular clouds (\citealt{Hollenbach09}). The latter are detected in the submm (e.g. \citealt{Neufeld00,Neufeld02,Plume04}), and are not included in this model. As a consequence the fitted \textit{cold extended model} for reproducing the relatively low-J lines of the CO band is overestimating the $\sim$R(4) to R(6) transitions. Residuals are also present in high-J transition of the $^{13}$CO $\nu$=1-0 and C$^{18}$O $\nu$=1-0 bands, suggesting that the $^{12}$CO/$^{13}$CO (and $^{12}$CO/C$^{18}$O) ratios is higher.

\section{Best-fit torus model images}
\label{torusimages}
We find a high optical depth ($\tau_{9.7\mu m}\sim$14.5) and highly inclined torus ($\sim$80 deg) with a half-opening angle of $\sim$30 deg from the best-fit torus model.  Following the same method as in \citet{Efstathiou95}, we obtain multi-wavelength images of the best-fit torus model (see Fig. \ref{torussimulaed}). Finally, by fitting a Gaussian, we measured sizes of $\sim$3 and 11\,pc (diameter) at 5\,$\mu$m and 1\,mm, respectively.

\begin{figure}
\centering
\par{
\includegraphics[width=4.2cm]{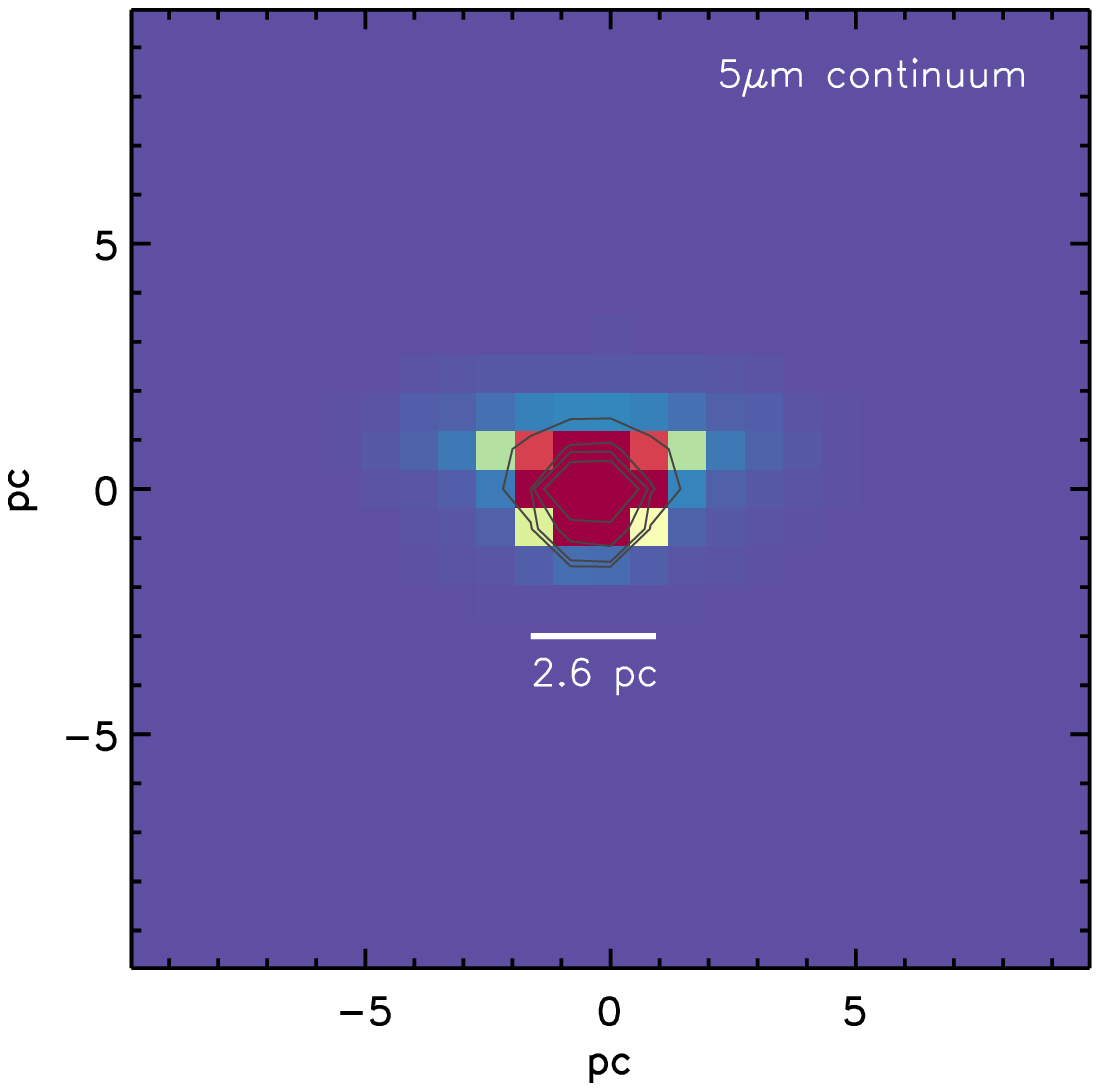}
\includegraphics[width=4.2cm]{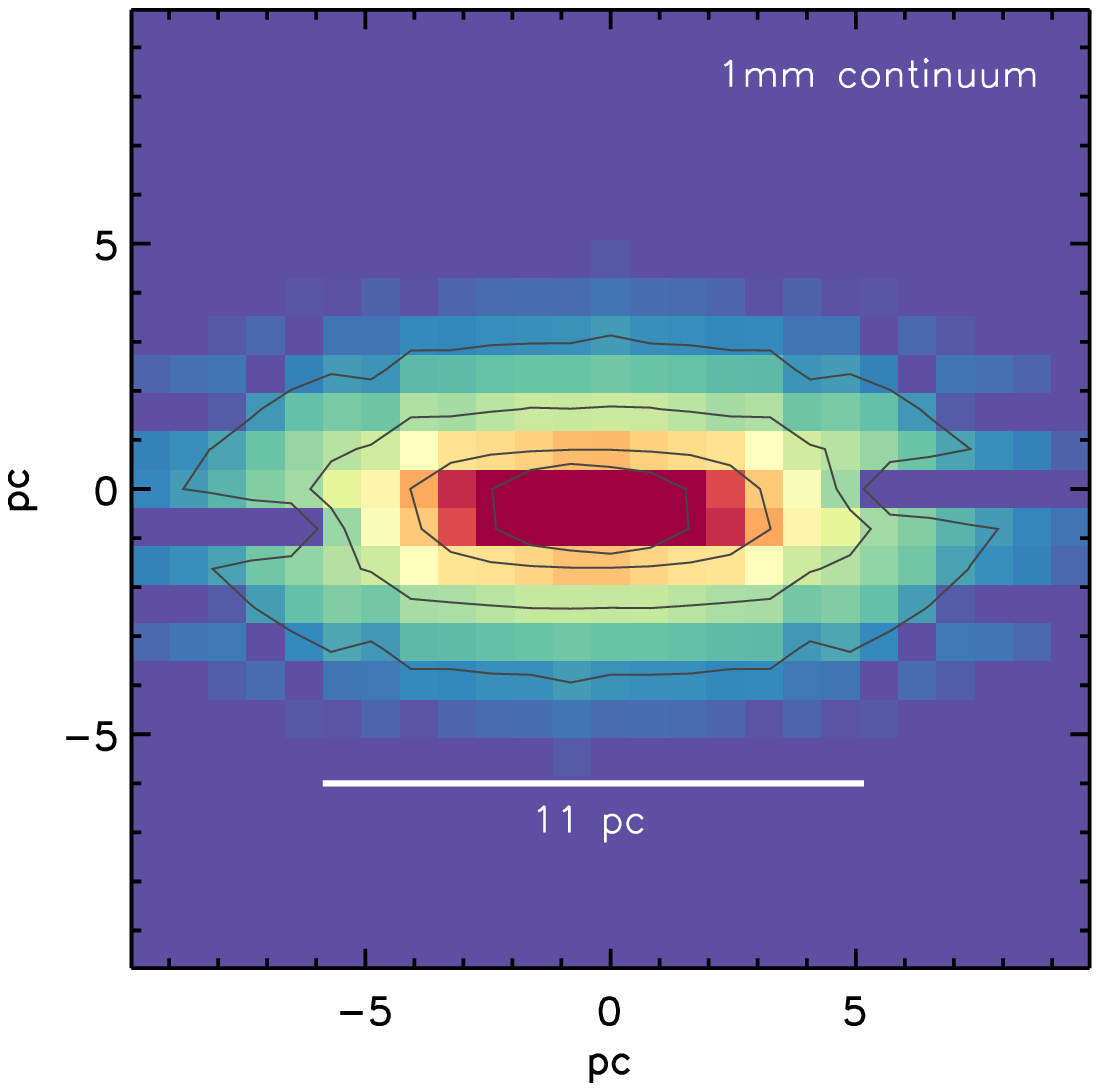}
\par}
\caption{Intensity map of the best-fit torus model at 5\,$\mu$m (left) and 1\,mm (right).}
\label{torussimulaed}
\end{figure}

\section{Metallicity}
\label{metallicity}
\begin{table}
\caption{Ar, Ne, and H extinction corrected fluxes. Ellipsis indicates undetected transitions.}
\label{tbl_metals}
\centering
\begin{small}
\begin{tabular}{lcccccc}
\hline \hline
\\
Transition & $\lambda$ & Flux \\
& ($\mu$m) &  (10$^{-15}$\,erg\,cm$^{-2}$\,s) \\
\hline
{[\ion{Ar}{ii}]} & 6.99  & 16.67 $\pm$ 0.77\\
{[\ion{Ar}{iii}]} & 8.99 & 20.97 $\pm$ 0.90\\
{[\ion{Ar}{v}]} & 7.90 & $\cdots$ \\
{[\ion{Ar}{v}]} & 13.10 & $\cdots$ \\
{[\ion{Ar}{vi}]} & 4.53 & $\cdots$ \\
{[\ion{Ne}{ii}]} & 12.81  & 98.93 $\pm$ 5.13\\
{[\ion{Ne}{iii}]} & 15.56 & 81.26 $\pm$ 4.31\\
{[\ion{Ne}{v}]} & 14.32 & $\cdots$ \\
{[\ion{Ne}{v}]} & 24.32 & $\cdots$ \\
{[\ion{Ne}{vi}]} & 7.65 & $\cdots$ \\
\ion{H}{i} 6--5 & 7.46 & 5.14 $\pm$ 1.16 \\
\hline
\end{tabular}
\end{small}
\end{table}

We estimated the gas-phase metallicity of D1 using the [\ion{Ar}{ii}]6.99\,$\mu$m, [\ion{Ar}{iii}]8.99\,$\mu$m, [\ion{Ne}{ii}]12.81\,$\mu$m, [\ion{Ne}{iii}]15.56\,$\mu$m, and \ion{H}{i}  6--5~7.46\,$\mu$m~(Pf$\alpha$) emission lines (Table~\ref{tbl_metals}). We followed the prescription described in \citet{Verma03} and used PyNeb  v1.1.18 \citep{Luridiana2015} to derive the emissivities of the Ar, Ne, and H lines.

For SBs, the ionisation correction factor for Ar is small ($<$30\%) based photoionisation models \citep{MartinHernandez2002}. The effect on the Ne abundance is negligible. In addition, no emission from highly ionised Ar and Ne (e.g. \citealt{Bernete17,Pereira2017FIR}) related to the AGN (deeply obscured) is  detected (see Table~\ref{tbl_metals}). Therefore, we assumed that the ionic abundances derived from the [\ion{Ar}{ii}], [\ion{Ar}{iii}], [\ion{Ne}{ii}] and [\ion{Ne}{iii}] transitions trace the great majority of Ar and Ne (i.e. Ar\slash H $\simeq$ (Ar$^+$ + Ar$^{++}$)\slash H$^+$ and Ne\slash H $\simeq$ (Ne$^+$ + Ne$^{++}$)\slash H$^+$), respectively.

For an electron temperature of 10000\,K (e.g. \citealt{Bernard-Salas2009}), the derived Ar and Ne abundance corresponds to 0.62--0.79 times the solar value (assuming the solar \hbox{Ar\slash H} and \hbox{Ne\slash H} derived in \citealt{Asplund21}).  This metallicity is lower than the range generally observed in U\slash LIRGs (\citealt{Rupke2008, Pereira2017FIR, Chartab2022}). 

We also derived the dust mass of D1 (2.3$\times$10$^6$\,M$_\sun$) from the grey body fit in Section \ref{modelling_torus} (following the same method as in \citealt{Klaas93}). Using the ALMA CO\,(3--2) flux reported by \citet{Wu22}, the CO\,(3--2)/CO\,(1--0) ratio (measured in K\,km\,s$^{-1}$) of 0.5-1 and the conversion factor generally used for ultraluminous IR galaxies ($\alpha_{\rm CO}$=0.8~M$_{\sun}$[K~km~s$^{-1}$~pc$^2$]$^{-1}$; for instance, \citealt{Bolatto13}), we obtained a {\textit{M}}(H$_2$) of $\sim$1.6-3.1$\times$10$^8$\,M$_\sun$ and a hydrogen column density of log N$_{\rm H}$ (cm$^{-2})\sim24.1-24.4$ in a beam of $\sim$134\,pc$\times$107\,pc for D1. \citet{Leech10} reported a CO\,(3--2)/CO\,(1--0) ratio of $\sim$0.5 from integrated values of luminous IR galaxies, but in dense and hot clumps, the CO(3-2) excitation might be higher; thus, we discuss the results using a CO\,(3--2)/CO\,(1--0) ratio ranging from 0.5 to 1. We also derived a high excitation temperature (T$_{\rm ex}$ $\sim$55\,K) for the CO (3-2) in D1, which is consistent with the presence of widespread PDRs within the beam (see also Section \ref{soda_modelling}) and intense emission of the ionised 7.7~$\mu$m PAH band. Finally, we found a gas-to-dust ratio in D1 of $\sim$70-135. Given the estimated metallicity of D1, its gas-to-dust ratio is relatively higher than that of "normal" SF galaxies (\citealt{remy14}). This suggests that a hard radiation field is present in D1, which might be affecting the balance between dust formation and destruction (\citealt{remy14}) and, thus, its dust properties.

\section{High IP emission lines}
\label{lines_appendix}
\begin{figure}
\centering
\par{
\includegraphics[width=5.5cm]{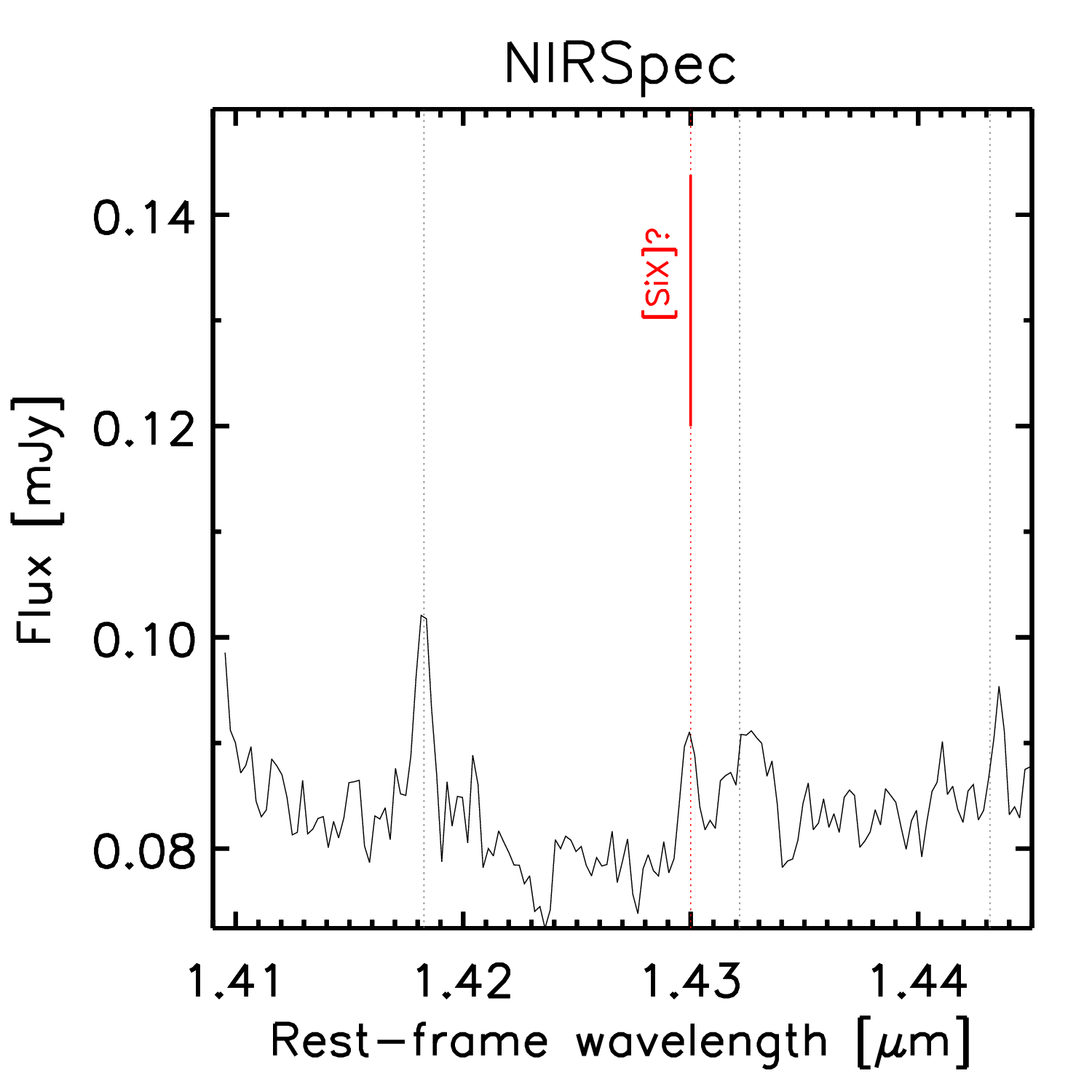}
\includegraphics[width=5.5cm]{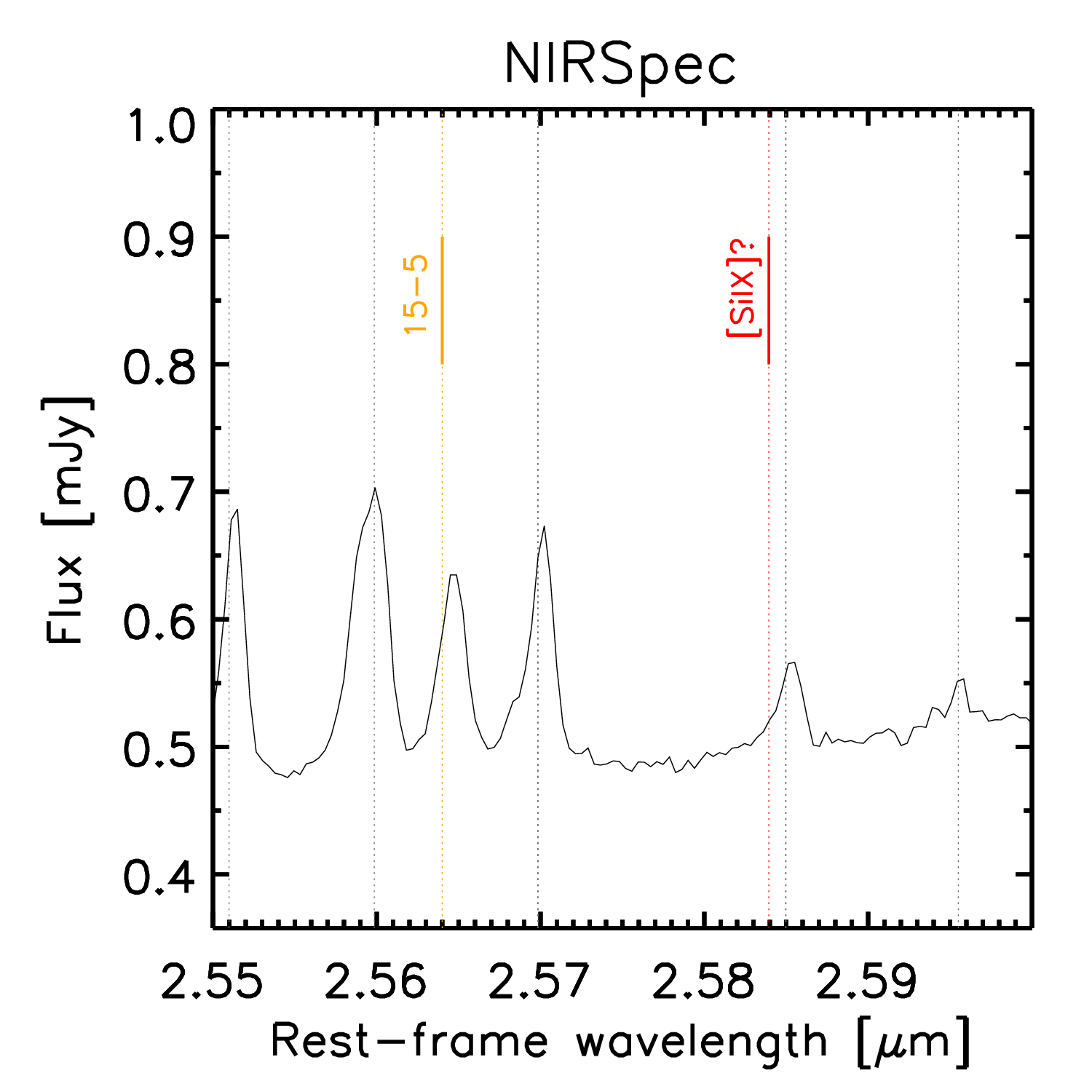}
\includegraphics[width=5.5cm]{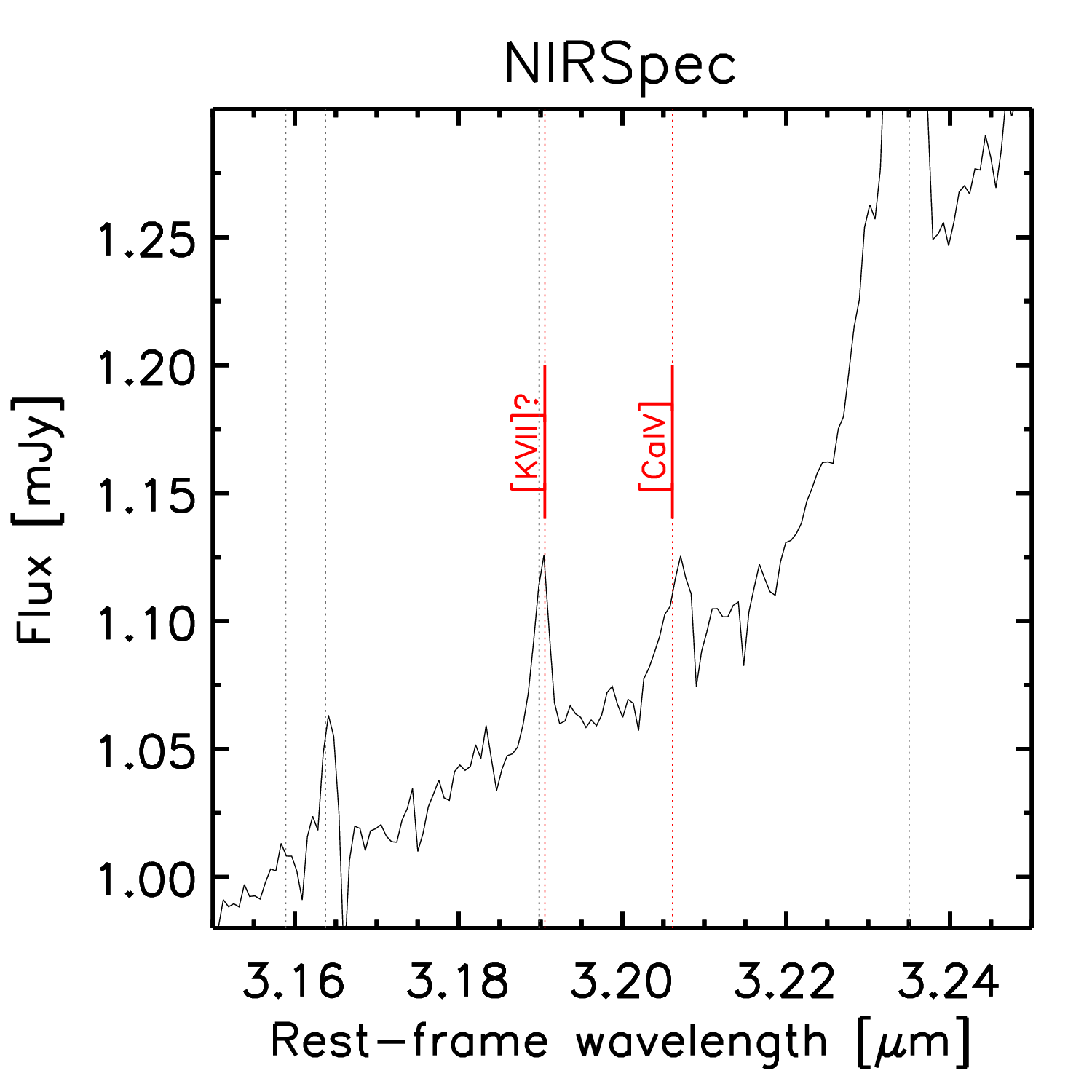}
\includegraphics[width=5.5cm]{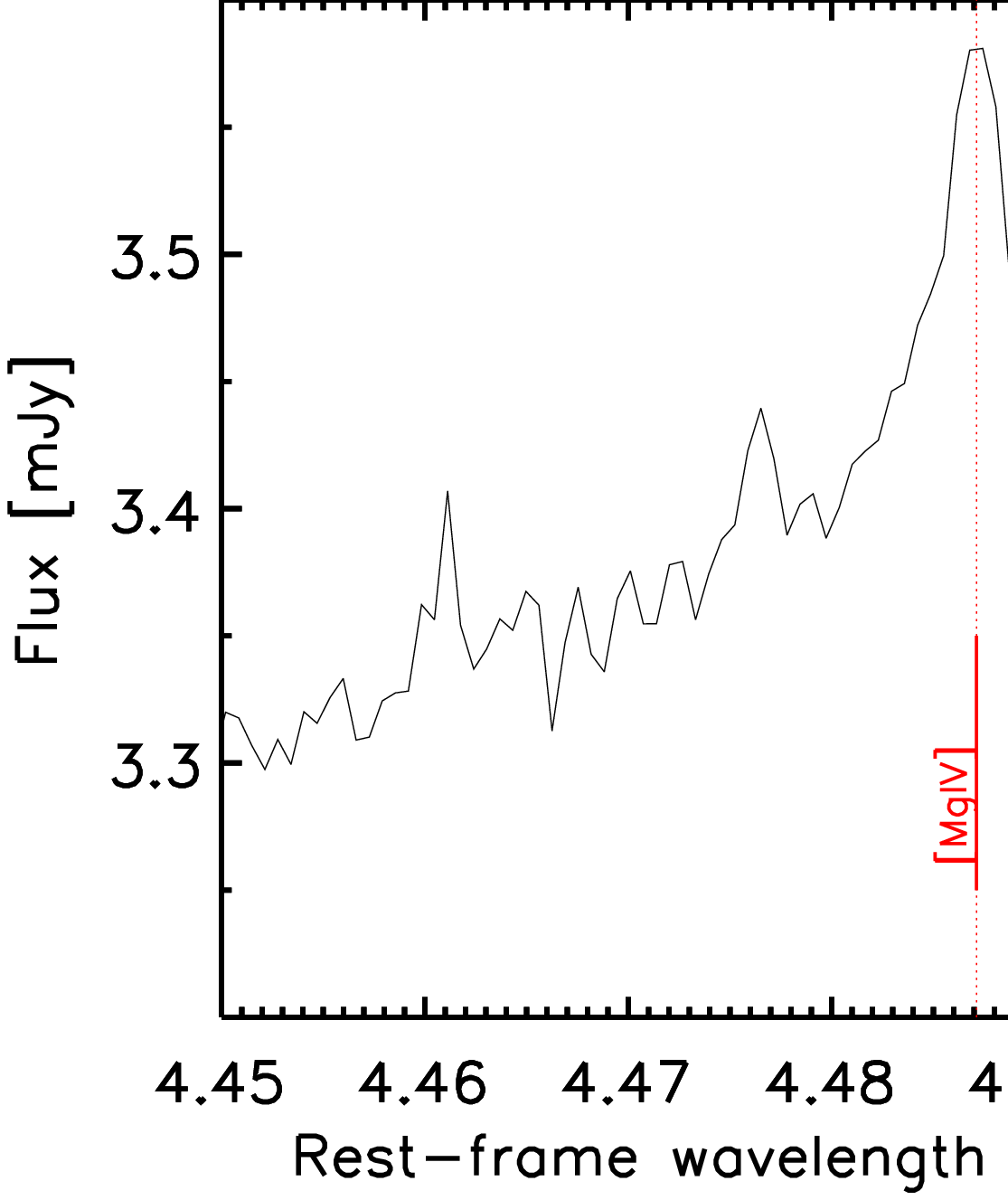}
\par}
\caption{Potentially detected high IP emission lines in II\,Zw\,096-D1. Black dotted, orange and red solid lines correspond to H$_2$, H recombination and high ionisation potential lines, respectively.}
\label{lines1}
\end{figure}

Figure \ref{lines1} shows the various high IP lines present in II\,Zw\,096-D1 with SNR greater than 3. Question marks indicate those that are not clearly detected. 

\end{appendix}

\end{document}